\documentclass[letterpaper]{llncs}  
\usepackage[utf8]{inputenc}
\usepackage{amsmath,amssymb,color,graphicx}
\usepackage[usenames,dvipsnames,svgnames,table]{xcolor}
\usepackage{blkarray}
\usepackage{verbatim}
\usepackage{cite}

\usepackage{caption}
\usepackage{subcaption}
\usepackage{hyperref}
\usepackage{wrapfig}

\usepackage{color,soul}

\newcommand{\remove}[1]{{}}

\usepackage[misc]{ifsym}

\usepackage{todonotes}

\title{Perception of Symmetries in Drawings of Graphs}
\author{Felice De Luca(\Letter)\inst{1} \and Stephen Kobourov\inst{1} \and Helen Purchase\inst{2}}
\institute{University of Arizona, USA\\\email{felicedeluca@email.arizona.edu, kobourov@cs.arizona.edu}
\and 
University of Glasgow, UK\\\email{helen.purchase@glasgow.ac.uk}
 }

\begin{document}

\maketitle

\begin{abstract}
Symmetry is an important factor in human perception in general, as well as in the visualization of graphs in particular. There are three main types of symmetry: reflective, translational, and rotational. We report the results of a human subjects experiment to determine what types of symmetries are more salient in drawings of graphs. 
We found statistically significant evidence that vertical reflective symmetry is the most dominant (when selecting among vertical reflective, horizontal reflective, and translational).
We also found statistically significant evidence that rotational symmetry is affected by the number of radial axes (the more, the better), with a notable exception at four axes.

\end{abstract}

\section{Introduction}
\begin{figure}
\centering
        \begin{subfigure}[b]{0.30\textwidth}
               \centering
 \includegraphics[height=2cm]{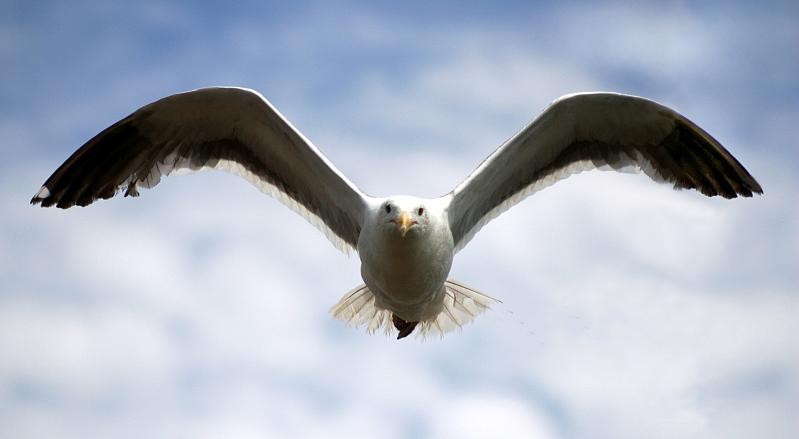}
                \caption{ }
                \label{fig:natversym}
        \end{subfigure}%
        \begin{subfigure}[b]{0.30\textwidth}
               \centering
 \includegraphics[height=2cm]{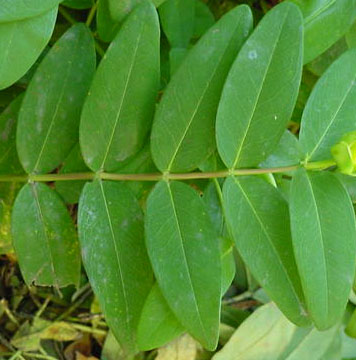}
                \caption{ }
                \label{fig:nattransym}
        \end{subfigure}%
        \begin{subfigure}[b]{0.30\textwidth}
               \centering
 \includegraphics[height=2cm]{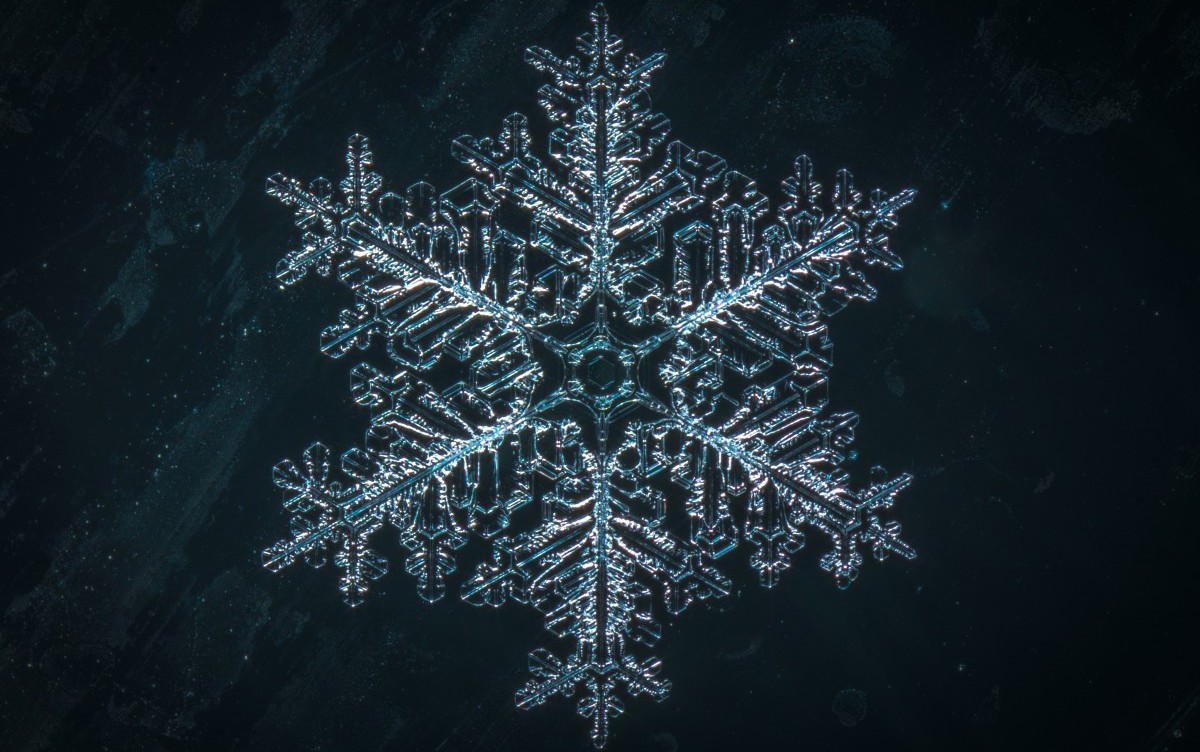}
                \caption{ }
                \label{fig:natradsym}
        \end{subfigure}%
        \caption{Symmetry in nature: (a) reflective, (b) tanslational, (c) rotational.}
        \label{fig:intro}
\end{figure}

Many objects in nature, from plants and animals to crystals and snowflakes, have symmetric patterns. Humans and other animals  have a nearly perfect reflective symmetry along a single axis; sea stars and snowflakes have repetitive patterns along two or more radial axes; leaves and flowers often have translational patterns of symmetry; see Fig.~\ref{fig:intro}. 

The perception of symmetry is one of the key concepts in Gestalt theory which studies how humans perceive different types of objects.
Symmetry has also been considered an important feature of well-drawn graphs, on the basis that depicting symmetries will reveal a graph's structure and properties~\cite{eades2013symmetric}. 
A natural question that arises is: which types of symmetry are easier to perceive and how does this affect drawings of graphs?
In this study we investigate this question, focusing on the reflective (also called ``mirror"), translational and rotational (also called ``radial") types of symmetries:
\begin{itemize}
    \item \textbf{Vertical}: A pattern is reflected across a vertical axis (\emph{reflective symmetry})
    \item \textbf{Horizontal}: A pattern is reflected across an horizontal axis (\emph{reflective symmetry with a 90 degree rotation})
    \item \textbf{Translational}: A pattern is repeated and shifted in the space
    \item \textbf{Rotational}: A pattern is repeated across radial axes with a given angle
\end{itemize}

We used synthetically generated graph drawings in a human subjects experiment to answer several questions related to the perception of symmetry. Specifically, we created 
graph layouts that exhibit different types of symmetries and asked our participants to select the more symmetric ones. 
We found statistically significant evidence that vertical reflective symmetry is the most dominant (when selecting among vertical, horizontal and translational).
We also found statistically significant evidence that  rotational symmetry is affected by the number of radial axes (the more, the better) with a notable exception at four axes.

\section{Related Work}
Gestalt theorists studied how objects are perceived, their view being that the perception of a whole object cannot be reduced to the sum of the perception of its parts~\cite{palmer1999theoretical}. They focused on describing the patterns we see in visual stimuli~\cite{bruce2003visual, ware2004information}, and in how we distinguish background from foreground~\cite{rock1990legacy}, devising a set of laws describing fundamental perceptual phenomena. Well-known Gestalt principles include {\em proximity} (things that are close together are perceived as being in a group), {\em similarity} (things that look ``similar'' are perceived as being in a group), {\em closure} (closed shapes are preferred to open shapes), and {\em symmetry}, which we discuss in more detail next.

Giannouli~\cite{giannouli2013visual} defines symmetries on the plane as ``transformations that preserve equal geometric distance," and identifies the three standard types:
translational, rotational and reflective symmetries.
Each of these have variations: the distance and direction of the transposition for translational symmetry, the number of times around a circle the object is repeated (known as its ``order") for rotational symmetry, and the angle of the axis (with horizontal and vertical being the most common) for reflective symmetry.

The human perception of symmetry has been studied in different contexts. For example, there is evidence that people remember figures as more symmetric and closer than they really are~\cite{tversky2005visuospatial} and that symmetry 
aids in the recall of abstract patterns~\cite{lai2009visual,schnore1967immediate}.

Several studies compare the effectiveness of different types of symmetry.
An early experiment by Corbalis and Roldan~\cite{corballis1974perception} compared vertical reflective symmetry with translation symmetry (limited to only horizontal translation).
They compared these two conditions in two forms -- where the two components were touching each other (i.e., creating a single holistic visual object) or where there was a horizontal distance between the two components (i.e., they were perceived as two separate objects). They concluded that vertical symmetry was more salient in the holistic case, but translation more salient when the two components are disjoint. Other early work~\cite{bruce1975violations} concluded that vertical symmetry is easier to detect than translational symmetry. 
Royer~\cite{royer1981detection} prioritizes horizontal and vertical reflective symmetry over diagonal reflection, with ``centric" (loosely comparable to rotational, despite Royer's stimuli being square in form) performing worse. The work by Palmer and Hemenway~\cite{palmer1978orientation} confirms the ordering: vertical reflective, horizontal reflective, diagonal reflective. 

Cattaneo {\em et al.}~\cite{cattaneo2017not} investigate the neurological basis for the perception of vertical and horizontal reflective symmetry and conclude that there is a ``partial" difference between the regions of the brain used in detecting these two symmetries.
Giannouli's~\cite{giannouli2013visual} review of research on the visual perception of symmetry 
finds that vertical reflective symmetry is more readily perceived than any other type. Similar findings are reported in an earlier review by Wageman~\cite{wagermans1995detection}, who ranks reflective symmetries as follows: vertical produces better recognition performance (faster or more accurate) than horizontal, which in turn performs better than diagonal.

Jennings and Kingdom~\cite{jennings2017searching} conducted experiments to compare the perception of different orders of rotational symmetry (3, 5 and 7), together with a vertical reflection condition. 
They conclude that it is easier to detect rotational symmetry as the order increases (as measured by response time), and that, in comparison with rotational symmetry, the vertical reflective symmetry condition performs better than $3$rd order, the same as $5$th order, and worse than $7$th order. 

Note that none of this work involved graphs or drawings of graph.

While some researchers have considered the application of Gestalt principles to graph drawing, such work is rather fragmented.
Wong and Sun~\cite{sun2005evaluating} create key criteria for the depiction of UML class diagrams based on the Gestalt theories, and then evaluate three UML diagram tools based on these criteria. 
Bennet {\em et al.}~\cite{bennett2007aesthetics} review the literature on graph drawing aesthetics with reference to Norman's~\cite{norman2004emotion} stages of perception (visceral, behavioral and reflective), including the Gestalt theories in the visceral stage. They conclude that more work needs to be done to validate the common graph drawing aesthetic criteria with respect to perceptual theories.
Nesbitt and Freidrich~\cite{nesbitt2002applying} discuss some of the Gestalt theories in relation to graphs that evolve over time, although symmetry is not explicitly considered. Lemon {\em et al.}'s experiments~\cite{lemon2007empirical} show that the principles of similarity, proximity and continuity affect the comprehension of complex software diagrams.
Rusu {\em et al.}~\cite{rusu2011using} focus on the principle of continuity, and proposed a method for reducing visual clutter created by edge crossings by creating gaps in the edges. This is embodied in the partial edge drawing algorithms of Bruckdorfer {\em et al.}~\cite{bruckdorfer2012mad} and Burch {\em et al.}~\cite{burch2011evaluating}.
Marriott {\em et al.}~\cite{marriott2012memorability} conduct an experiment looking at what features of small graph drawings made them most memorable. Their experimental conditions explicitly relate to the Gestalt principles of symmetry, continuity, orientation and proximity and their finding indicate that drawings that exhibit symmetry and continuity are amongst those most readily recalled.

Eades advocates the use of
algorithms that aim to draw graphs with ``as much symmetry as possible"~\cite{eades2013symmetric}. Early experiments investigating the relative importance of different graph drawing aesthetics find support for the depiction of symmetry in terms of performance on graph-reading tasks~\cite{purchase1995validating, purchase1997aesthetic}.
Two computational methods have been proposed for measuring the extent of symmetry in a graph drawing, a non-trivial task in and of itself. 
The method proposed by Purchase~\cite{purchase2002metrics} considers only reflective symmetry. It generates potential axes of symmetry between all pairs of vertices, and determines the existence of symmetric sub-graphs (edges reflected around the axis, with a tolerance) for each axis. Klapaukh's method~\cite{klapaukh2014empirical} uses an edge-based metric that includes rotational and translation symmetries in addition to vertical ones. Welch and Kobourov~\cite{welch2017measuring} studied which of these two algorithms best correlates with the human perception of symmetry, with results that suggest that a graph drawing with vertical symmetry is considered more symmetric than the identical drawing presented at a slightly different orientation, and that the greater the extent of symmetry in a drawing, the faster the participants' response. 

While 
there has been extensive experimental research in the perception literature comparing the different types of symmetry in a variety of artificial stimuli, no comparable work has been performed to investigate the perception of symmetry in graph drawings. We therefore extend the work done so far by conducting experiments that specifically considers which of the three types of symmetry are more salient in drawings of graphs, including several additional variations.

\section{Research Questions}
We investigate the perception of graph drawings that exhibit
three types of symmetry:
reflective (vertical and horizontal), translational and rotational. 
Rather than attempting to draw existing graphs that embody 
such symmetries (a difficult 
task), we create symmetric graph drawings by duplicating graph-substructures. Specifically, we draw a small graph, make a duplicate, place the duplicate(s) appropriately (according to the type of symmetry), and join the components together to create a graph drawing that exhibits the desired symmetry. 

Since rotational symmetry is visually very different from reflective and translational symmetry, we address two separate research questions:
\begin{enumerate}
\item What is the relative ranking of reflective and translational symmetries for drawings of graphs?
\item What is the impact of the number of axes (order) for rotational symmetry? 
\end{enumerate}

\subsection{The Symmetric Graph Drawings}
\begin{figure}
\centering
        \raisebox{.3\height}{\begin{subfigure}[b]{0.12\textwidth}
        \includegraphics[width=\linewidth]{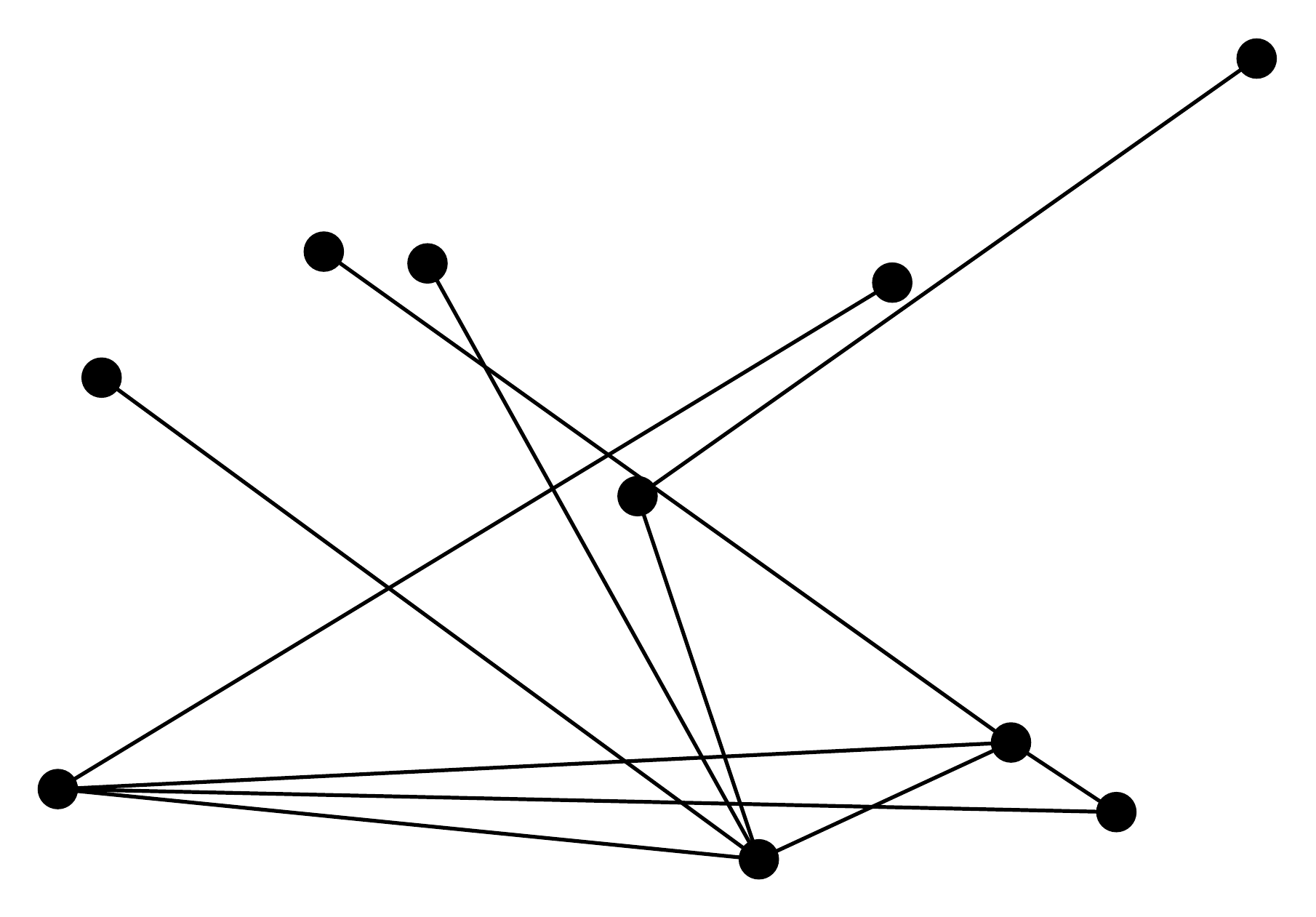}
                \caption{ }
                \label{fig:layout1}
        \end{subfigure}%
        }
        \begin{subfigure}[b]{0.09\textwidth}
    \includegraphics[width=\linewidth]{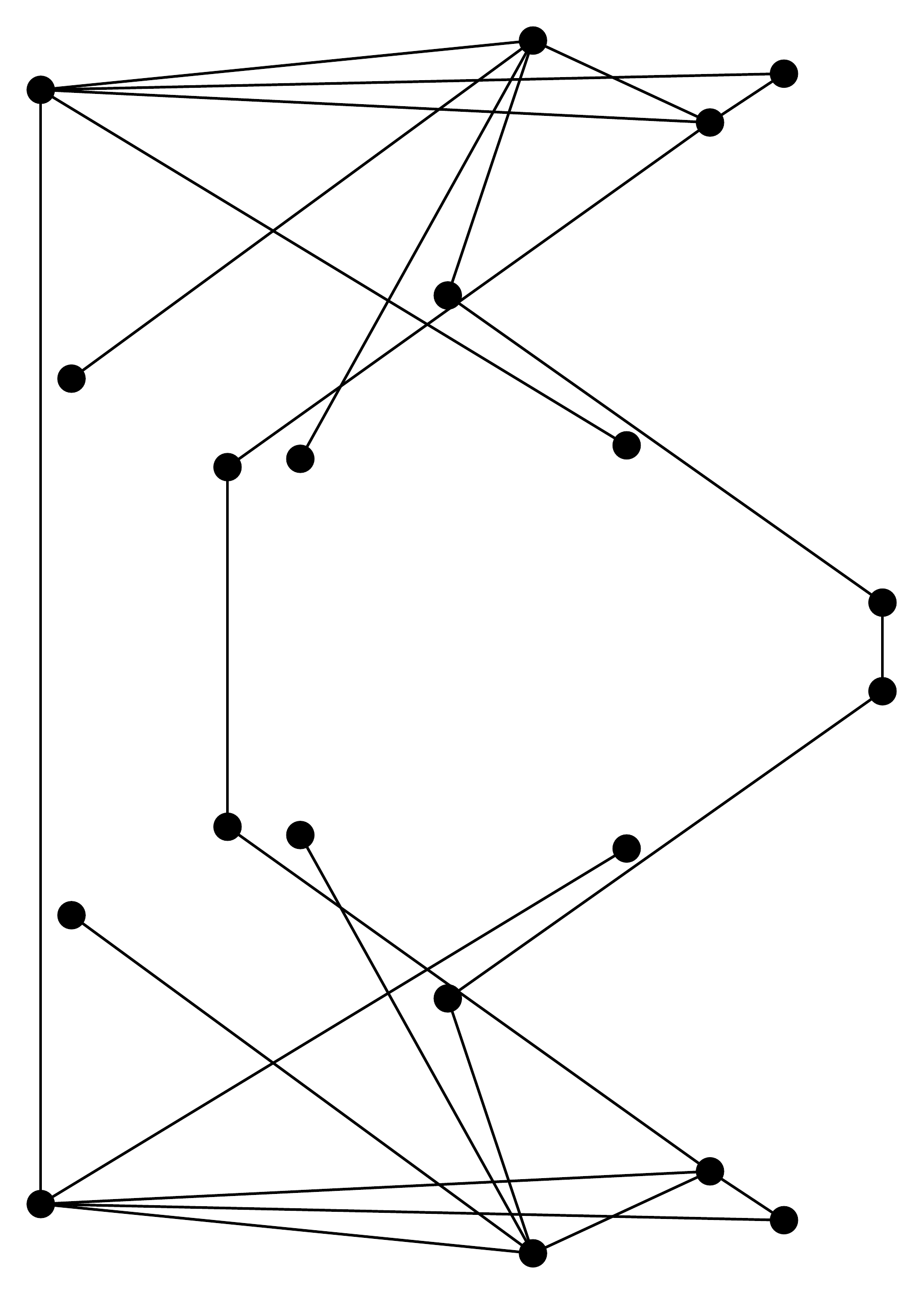}
                \caption{ }
                \label{fig:layout1h}
        \end{subfigure}%
         \raisebox{.1\height}{
    \begin{subfigure}[b]{0.10\textwidth}
    \includegraphics[width=\linewidth]{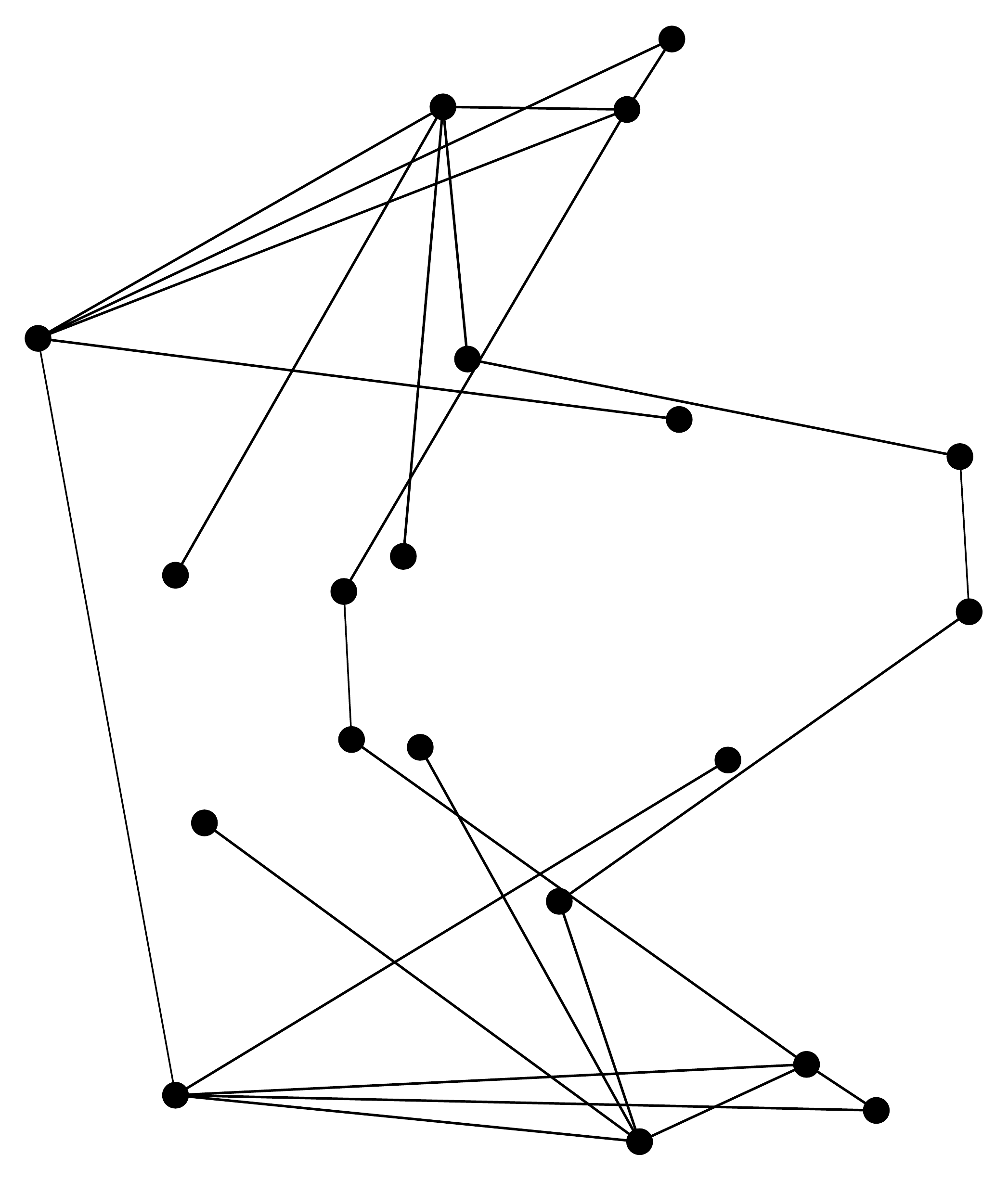}
                \caption{ }
                \label{fig:layout1hr}
        \end{subfigure}%
    }
        \raisebox{.3\height}{
        \begin{subfigure}[b]{0.12\textwidth}
        \includegraphics[width=\linewidth]{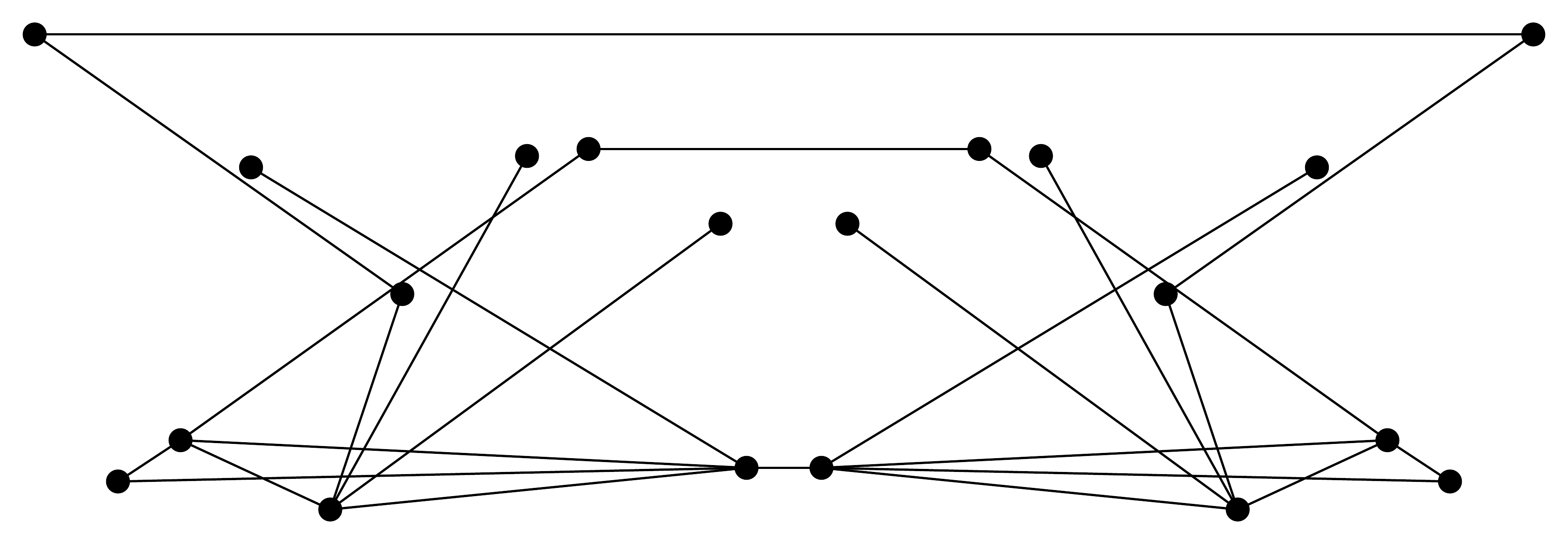}
                \caption{ }
                \label{fig:layout1v}
        \end{subfigure}%
        \begin{subfigure}[b]{0.12\textwidth} \includegraphics[width=\linewidth]{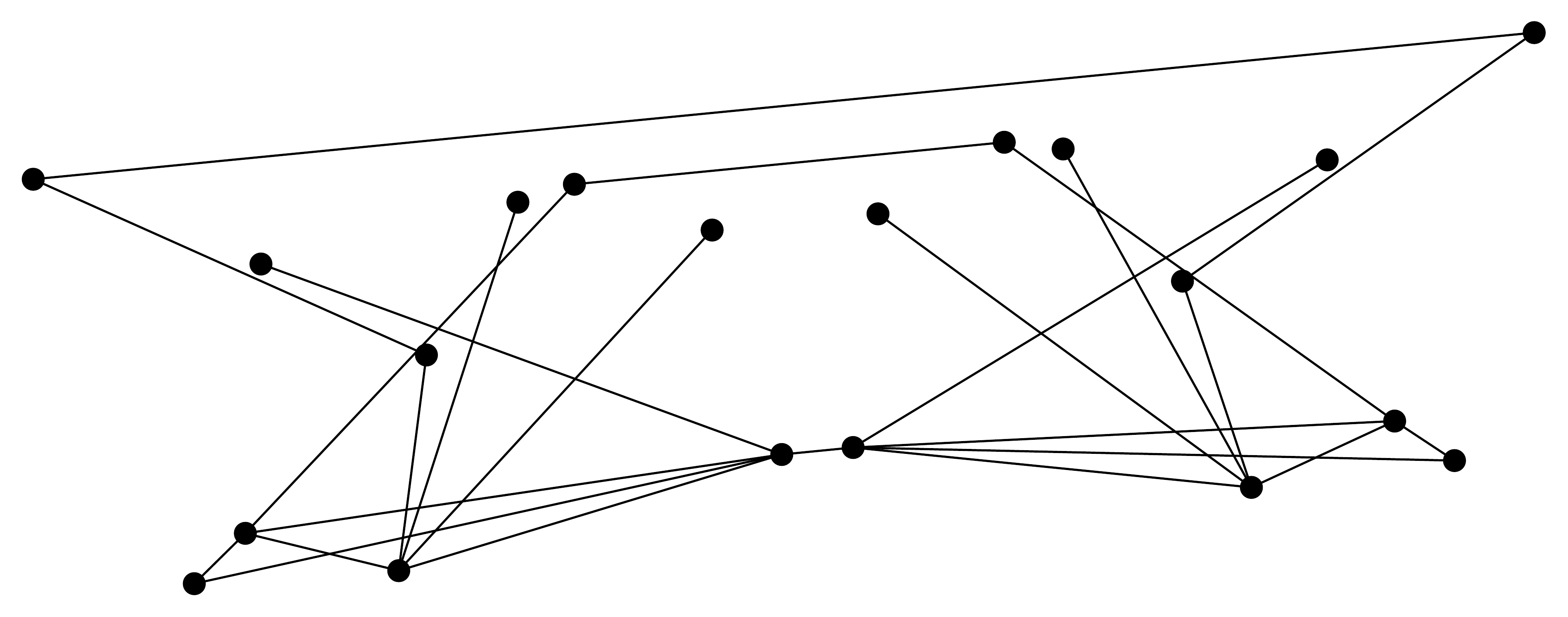}
                \caption{ }
                \label{fig:layout1vr}
        \end{subfigure}%
        
        \begin{subfigure}[b]{0.12\textwidth}
                \includegraphics[width=\linewidth]{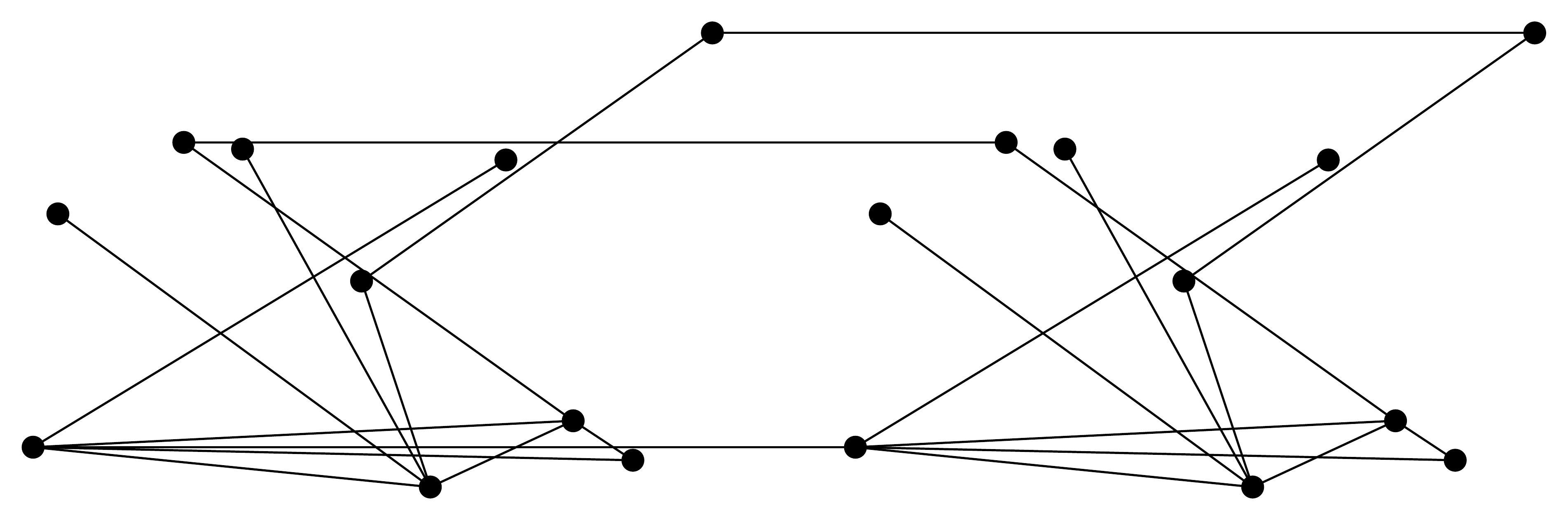}
                \caption{ }
                \label{fig:layout1t}
        \end{subfigure}%
        \begin{subfigure}[b]{0.12\textwidth}
                \includegraphics[width=\linewidth]{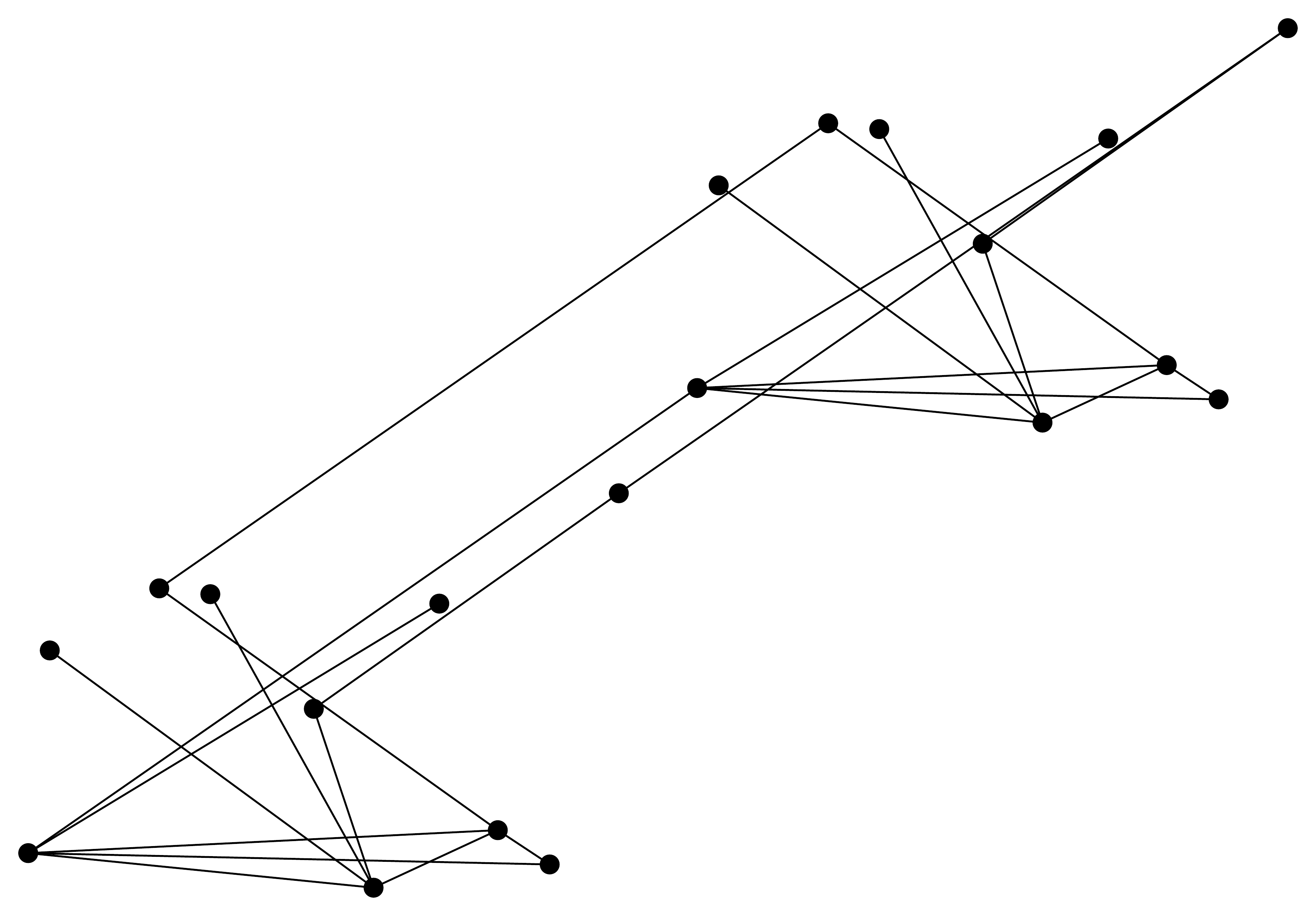}
                \caption{ }
                \label{fig:layout1tr}
        \end{subfigure}%
        \begin{subfigure}[b]{0.12\textwidth}
                \includegraphics[width=\linewidth, height=10mm, keepaspectratio]{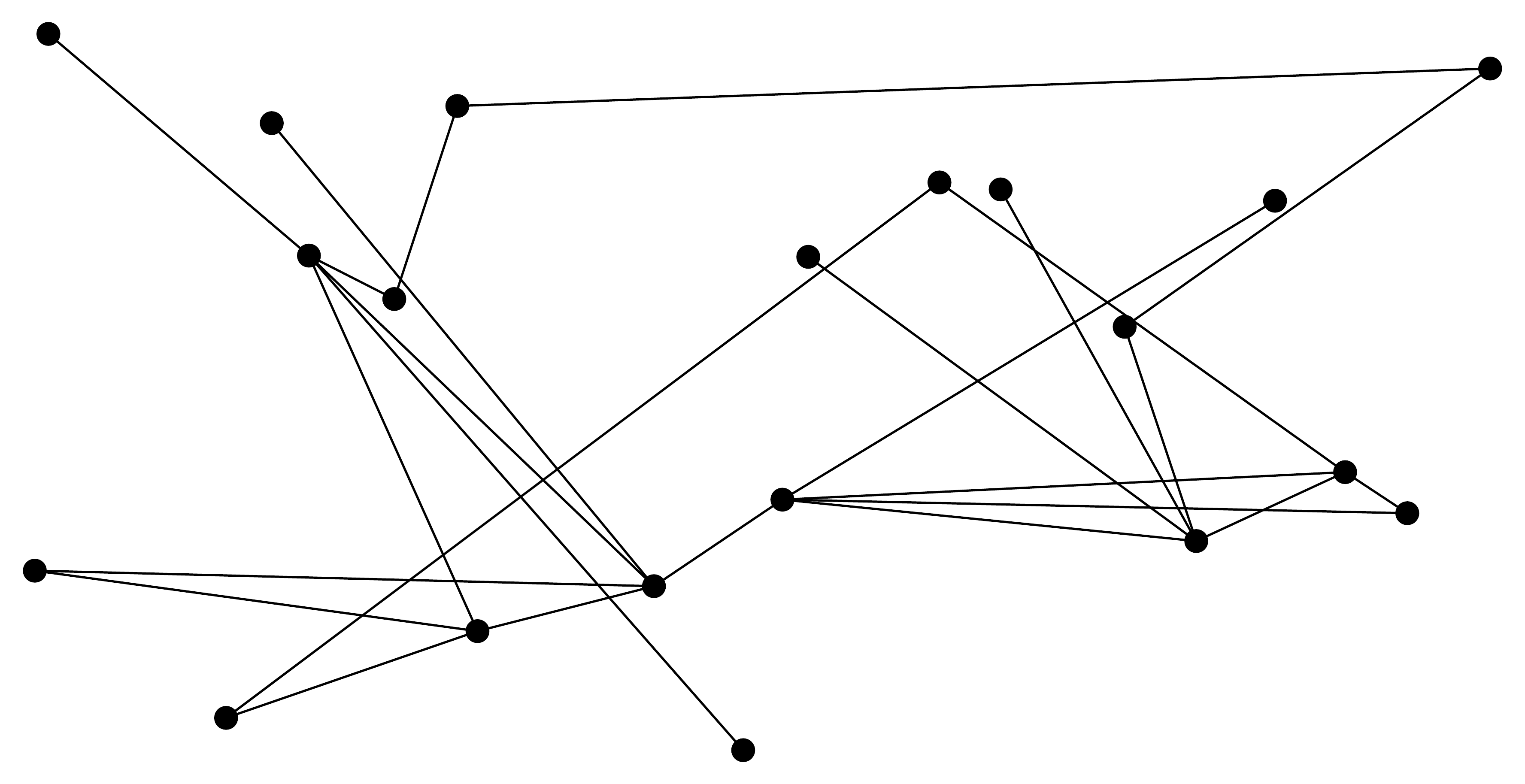}
                \caption{ }
                \label{fig:layout1ns}
        \end{subfigure}%
        }
        \caption{Example layouts in the dataset: (a) Base graph, and its (b) H, (c) Hr, (d) V, (e) Vr, (f) T, (g) Tr, and (h) NS.}\label{fig:layouts}
\end{figure}

Our experiment considers several different types of symmetry: horizontal (H), horizontal with rotation (Hr), vertical (V), vertical with rotation (Vr), translational (T), translational with rotation (Tr), rotational with fixed components (RC) and rotational with fixed vertices (RV). For a baseline, we also have a non-symmetric version (NS); see Fig.~\ref{fig:layouts}.
We consider two variants of rotational symmetry (RC and RV) in order to take into account the effect of the number of rotational axes and the effect of different graph sizes.

Ideally, all the stimuli should represent exactly the same graph, but since this would be impossible (especially for the rotational drawings), we attempt to impose some consistency by using the same ``base graph" from which the larger graphs are derived.

We create the horizontal, vertical and translational drawings (HVT, for short) as follows. Let  $G_b = (V_b , E_b)$ be a base graph drawn with a random layout such that each vertex $v_b \in V_b$ has positive coordinates; see Fig.~\ref{fig:layout1}.
Let $G_c = (V_c, E_c)$ be a copy of $G_b$ with the same layout. Then $G_s = (V_b \cup V_c, E_b \cup E_c \cup E)$ is created from the two graphs together with edge set $E$ connecting the vertices in $V_b$ to their copied version in $V_c$. We fix $|E| = 3$ and the vertices that are chosen for the connection are chosen at random. We use $G_s$ to create the layouts of the graphs in the HVT set by changing the coordinates of the vertices $v_c \in V_s$ as follows:

\begin{itemize}
    \item \textbf{H}: If $v_b = (x, y)$ then $v_c = (x, -y)$; see Fig.~\ref{fig:layout1h}.
    \item \textbf{Hr}: H version with a rotation with angle in $[0, 45]$; see Fig.~\ref{fig:layout1hr}.
    \item \textbf{V}: If $v_b = (x, y)$ then $v_c = (-x, y)$; see Fig.~\ref{fig:layout1v}.
    \item \textbf{Vr}: V version with a rotation with angle in $[0, 45]$; see Fig.~\ref{fig:layout1vr}.
    \item \textbf{T}: If $v = (x, y)$ then $v_c = (x-\delta, y)$ where $\delta$ is a shifting factor such that the bounding boxes of $V_b$ and $V_s$ do not overlap; see Fig.~\ref{fig:layout1t}.
    \item \textbf{Tr}: T version with a rotation with angle in $[0, 45]$; see Fig.~\ref{fig:layout1tr}.
    \item \textbf{NS}: Non symmetric (random) placement of the vertices in $V_s$; see Fig.~\ref{fig:layout1ns}.
\end{itemize}

We have two types of rotational drawings: maintaining the base graph component (the ``fixed component" version), and limiting the maximum number of vertices (the ``fixed vertices'' version). We use both methods in order to control for the possible confounding factor of different graph sizes in the first variant. 

\begin{figure}
\centering
        \begin{subfigure}[b]{0.14\textwidth}
                \includegraphics[width=\linewidth]{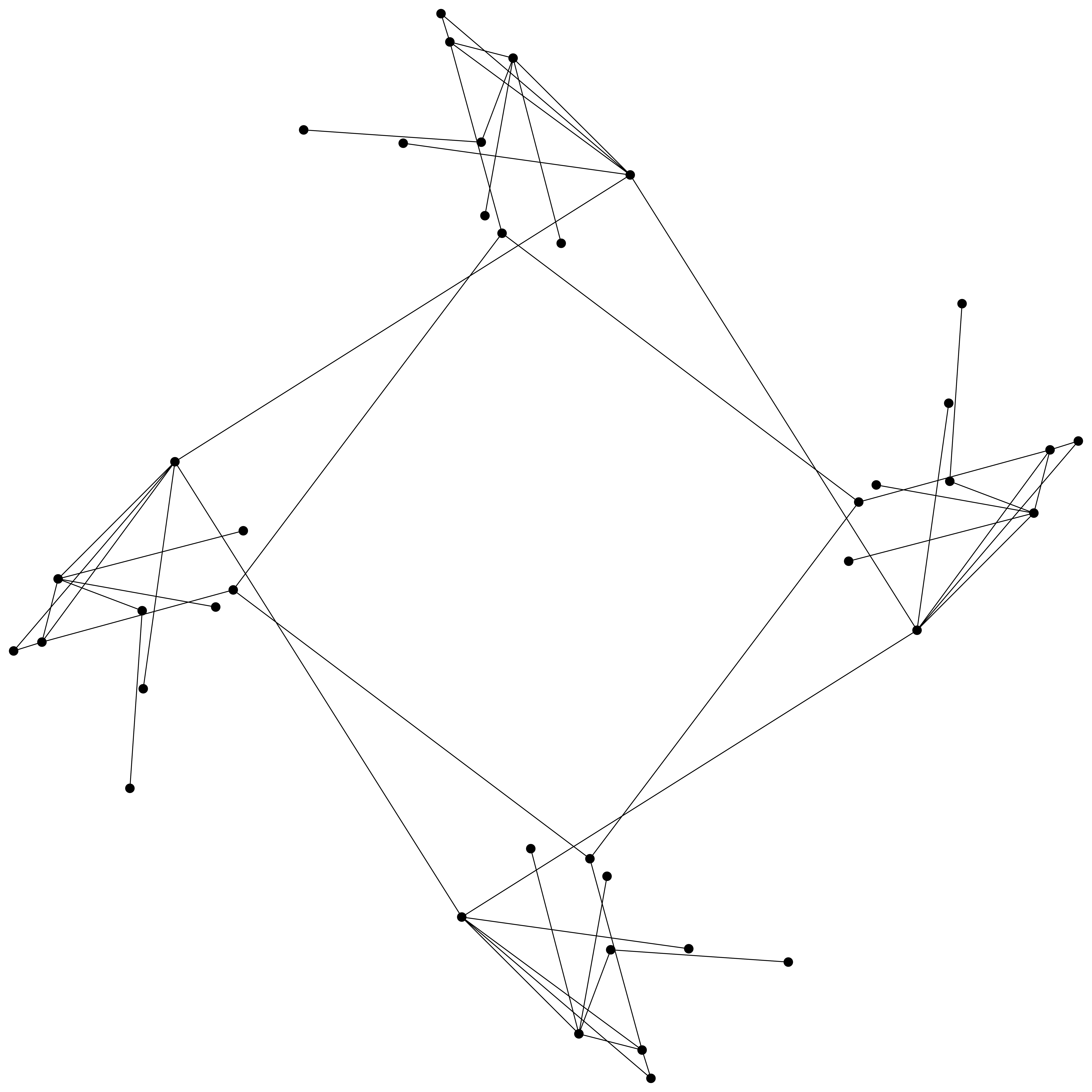}
                \caption{ }
                \label{fig:layoutRC4}
        \end{subfigure}%
        \begin{subfigure}[b]{0.14\textwidth}
                \includegraphics[width=\linewidth]{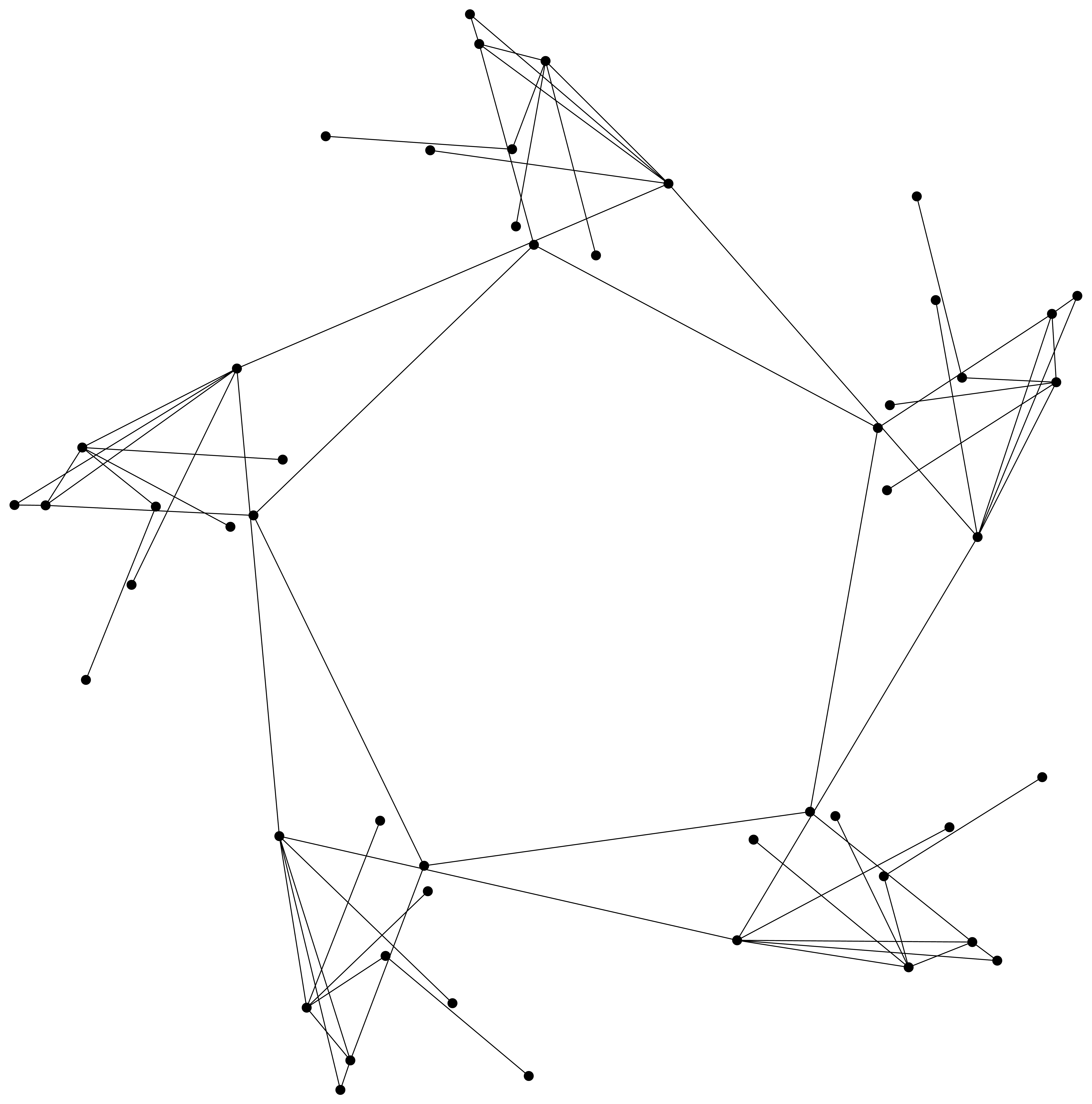}
                \caption{ }
                \label{fig:layoutRC5}
        \end{subfigure}%
        \begin{subfigure}[b]{0.14\textwidth}
                \includegraphics[width=\linewidth]{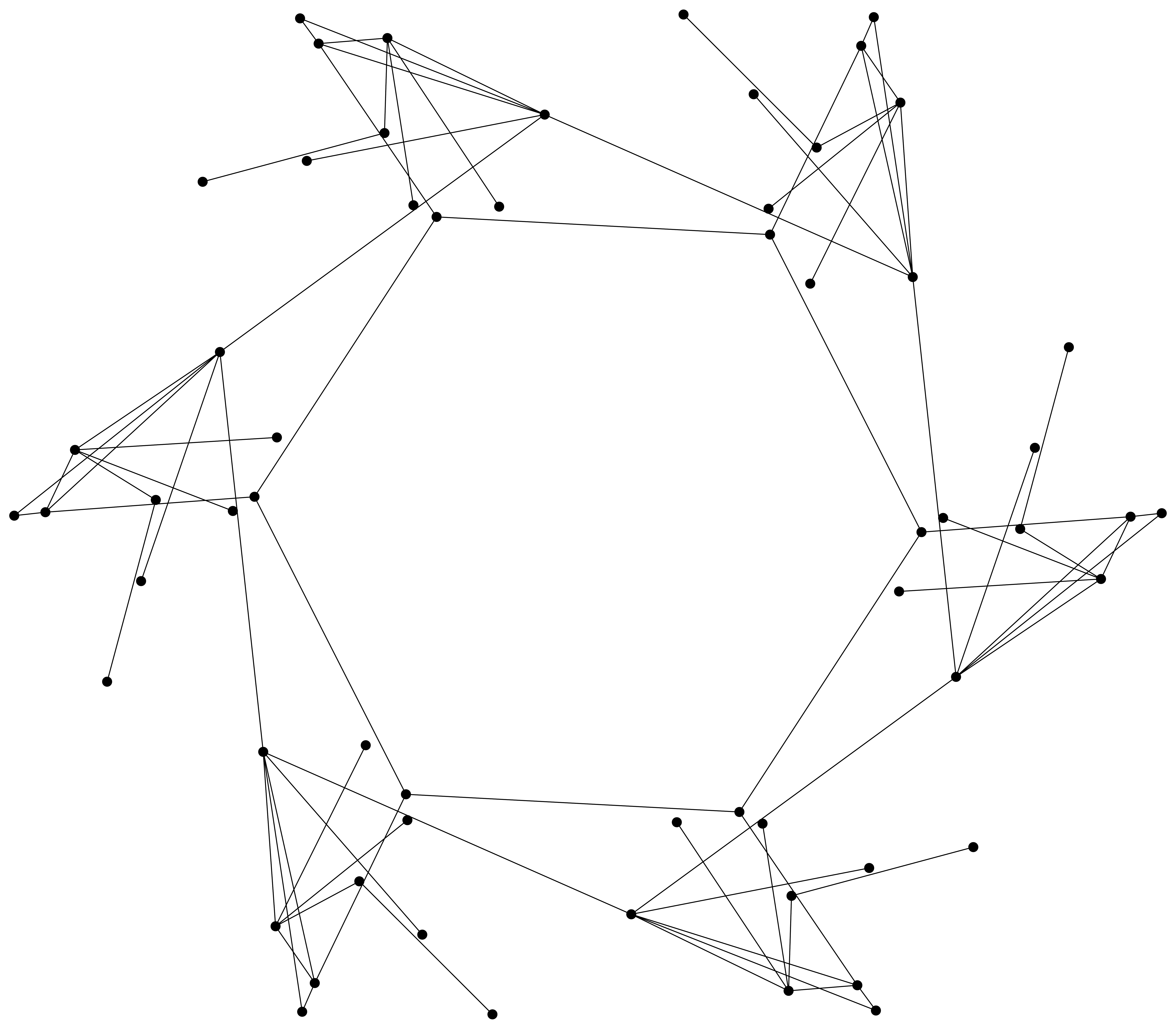}
                \caption{ }
                \label{fig:layoutRC6}
        \end{subfigure}%
        \begin{subfigure}[b]{0.14\textwidth}
                \includegraphics[width=\linewidth]{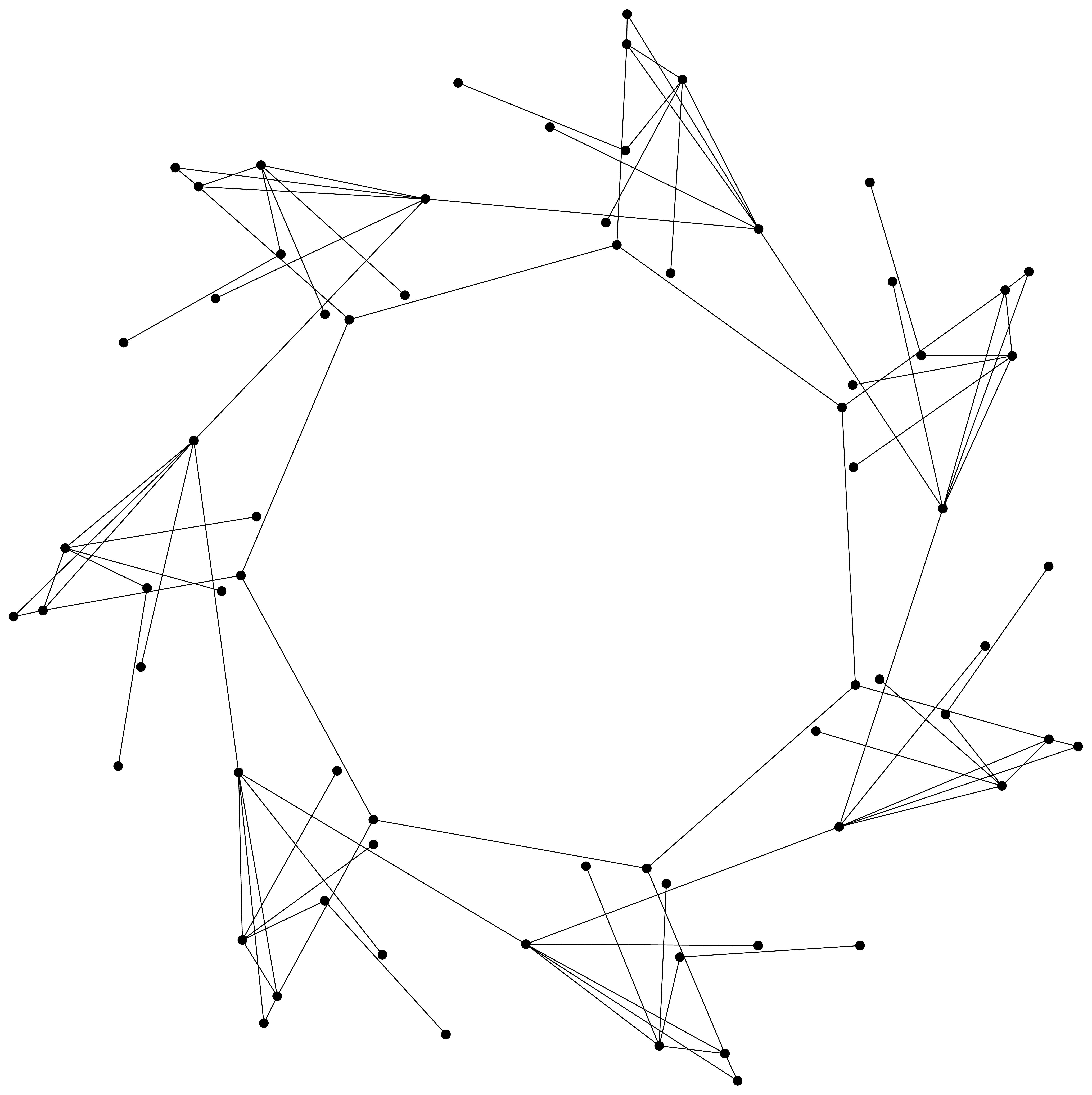}
                \caption{ }
                \label{fig:layoutRC7}
        \end{subfigure}%
        \begin{subfigure}[b]{0.14\textwidth}
                \includegraphics[width=\linewidth]{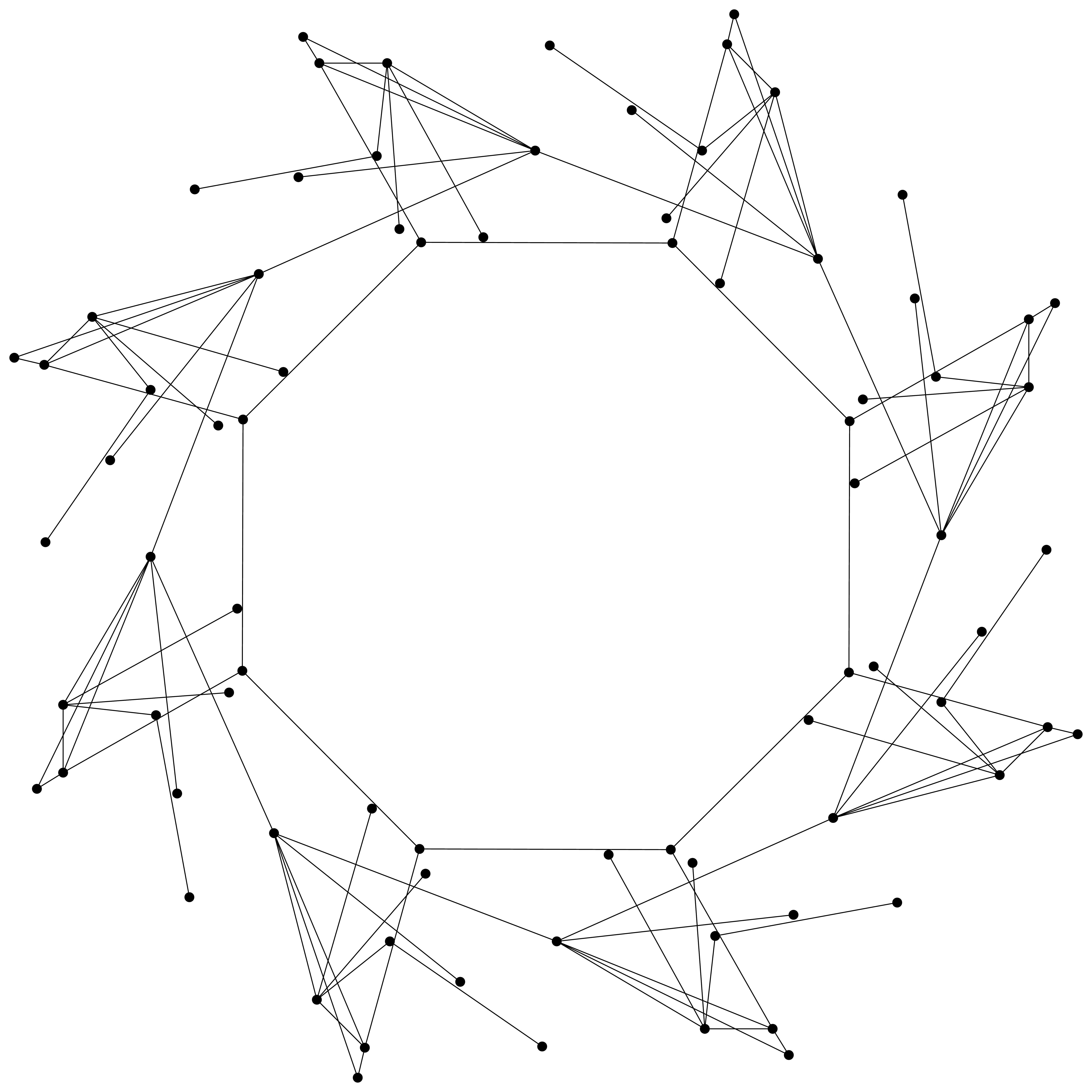}
                \caption{ }
                \label{fig:layoutRC8}
        \end{subfigure}%
        \begin{subfigure}[b]{0.14\textwidth}
                \includegraphics[width=\linewidth]{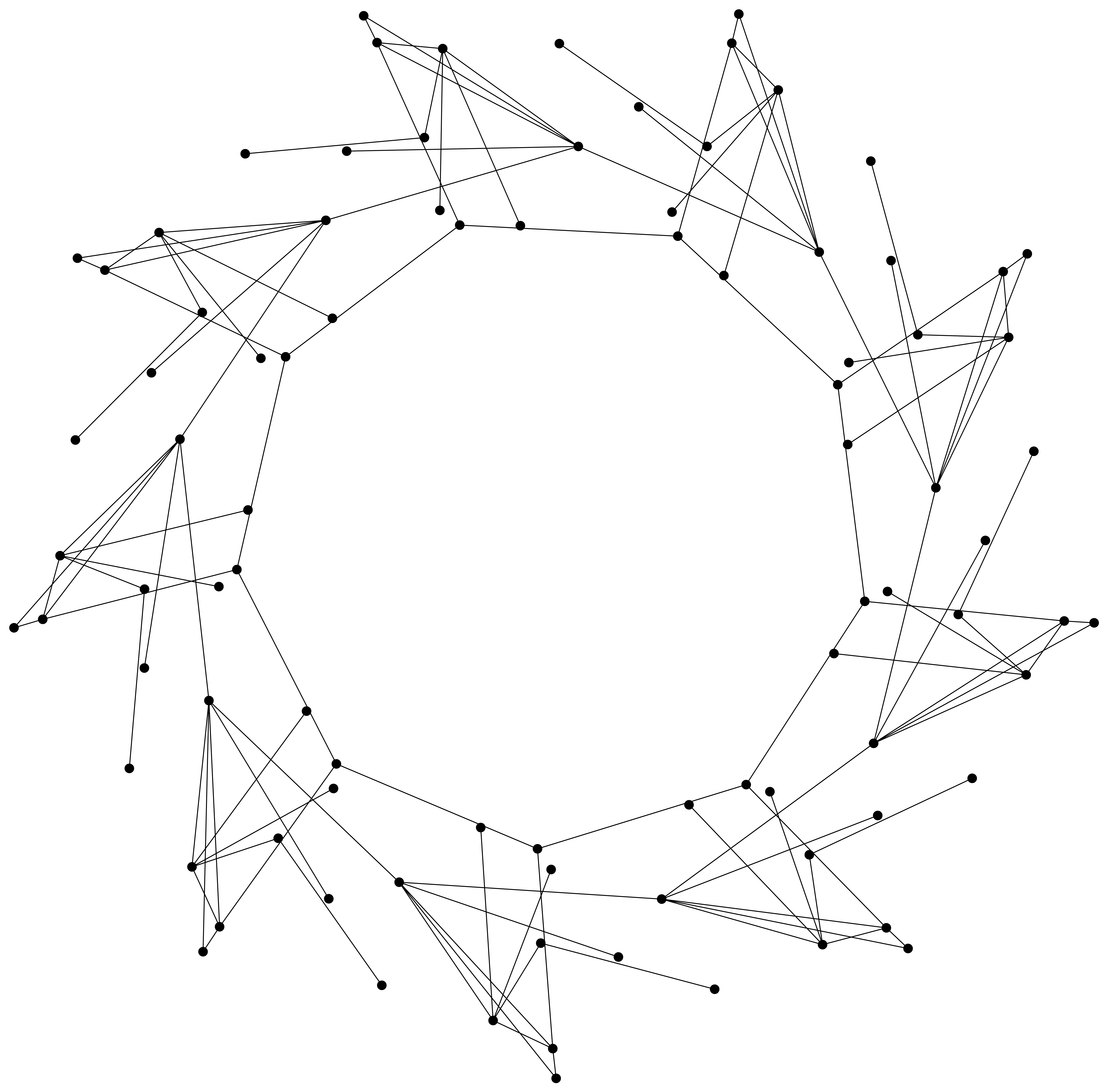}
                \caption{ }
                \label{fig:layoutRC9}
        \end{subfigure}%
        \begin{subfigure}[b]{0.14\textwidth}
                \includegraphics[width=\linewidth]{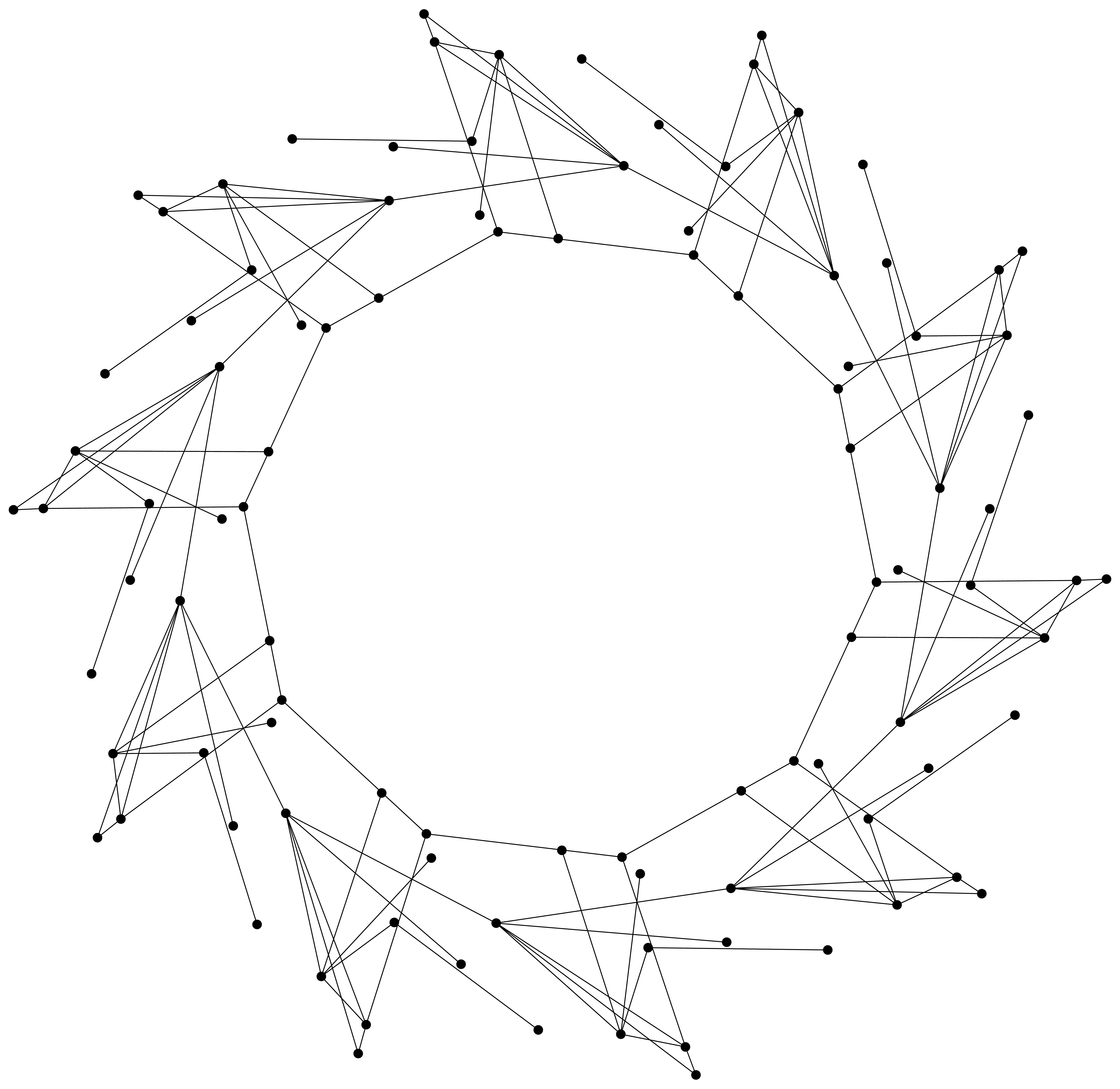}
                \caption{ }
                \label{fig:layoutRC10}
        \end{subfigure}%
        \caption{Example of rotational layouts with fixed components: (a) RC4, (b) RC5, (c) RC6, (d) RC7, (e) RC8, (f) RC9, and (g) RC10.}\label{fig:layoutsradC}
\end{figure}

The rotational fixed component (RC for short) versions are  symmetric layouts that repeat the base graph drawing around each axis. The base graph $G_b = (V_b, E_b)$ is a component of the layout of the symmetric graph $G_s$ and we create the different graphs as follows:
\begin{itemize}
    \item \textbf{RC\{X\}}: The drawing of each component is replicated on each of the $X = [4, \dots, 10]$ axes of symmetry. 
    By choosing two random vertices from a component, we connect each pair of rotationally consecutive components with edges to the corresponding vertices; 
    see Fig.~\ref{fig:layoutsradC}.
\end{itemize}

\begin{wraptable}[12]{r}{2.7cm}
	\small
	\begin{center}
				\caption{RV sizes}
		\begin{tabular}{ | l | c | r |}
			\hline
			& {V} & avg {E}\\ \hline
			\textbf{RV4} & 40 & 52 \\ \hline
			\textbf{RV5} & 50 & 65 \\ \hline
			\textbf{RV6} & 48 & 60.3\\ \hline
			\textbf{RV7} & 49 & 60.9\\ \hline
			\textbf{RV8} & 48 & 59.6\\ \hline
			\textbf{RV9} & 45 & 55.8\\ \hline
			\textbf{RV10} & 50 & 62\\ \hline
		\end{tabular}

		\label{tab:fixvertices}
	\end{center}
\end{wraptable}

\begin{figure}
\centering
        \begin{subfigure}[b]{0.14\textwidth}
                \includegraphics[width=\linewidth]{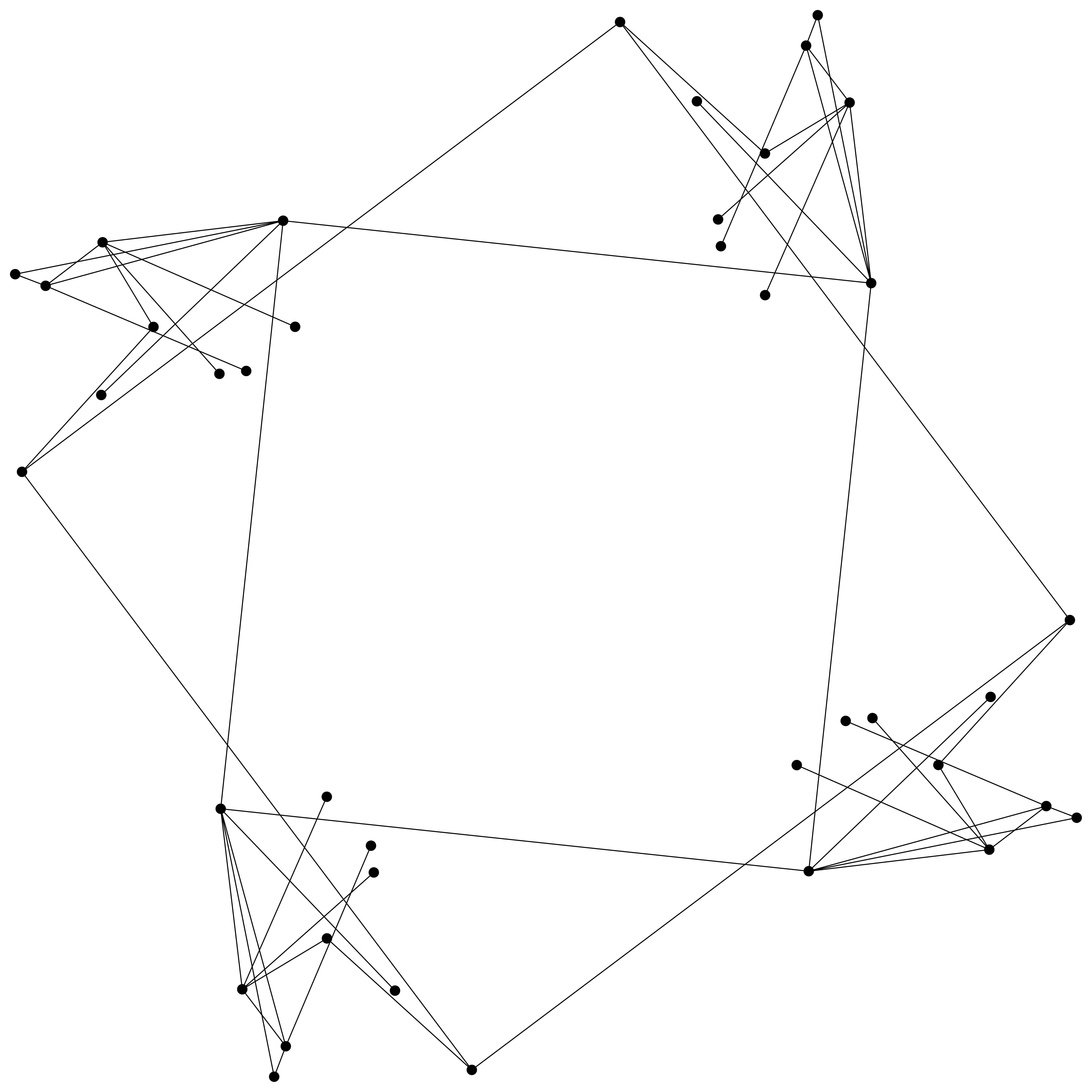}
                \caption{ }
                \label{fig:layoutRV4}
        \end{subfigure}%
        \begin{subfigure}[b]{0.14\textwidth}
                \includegraphics[width=\linewidth]{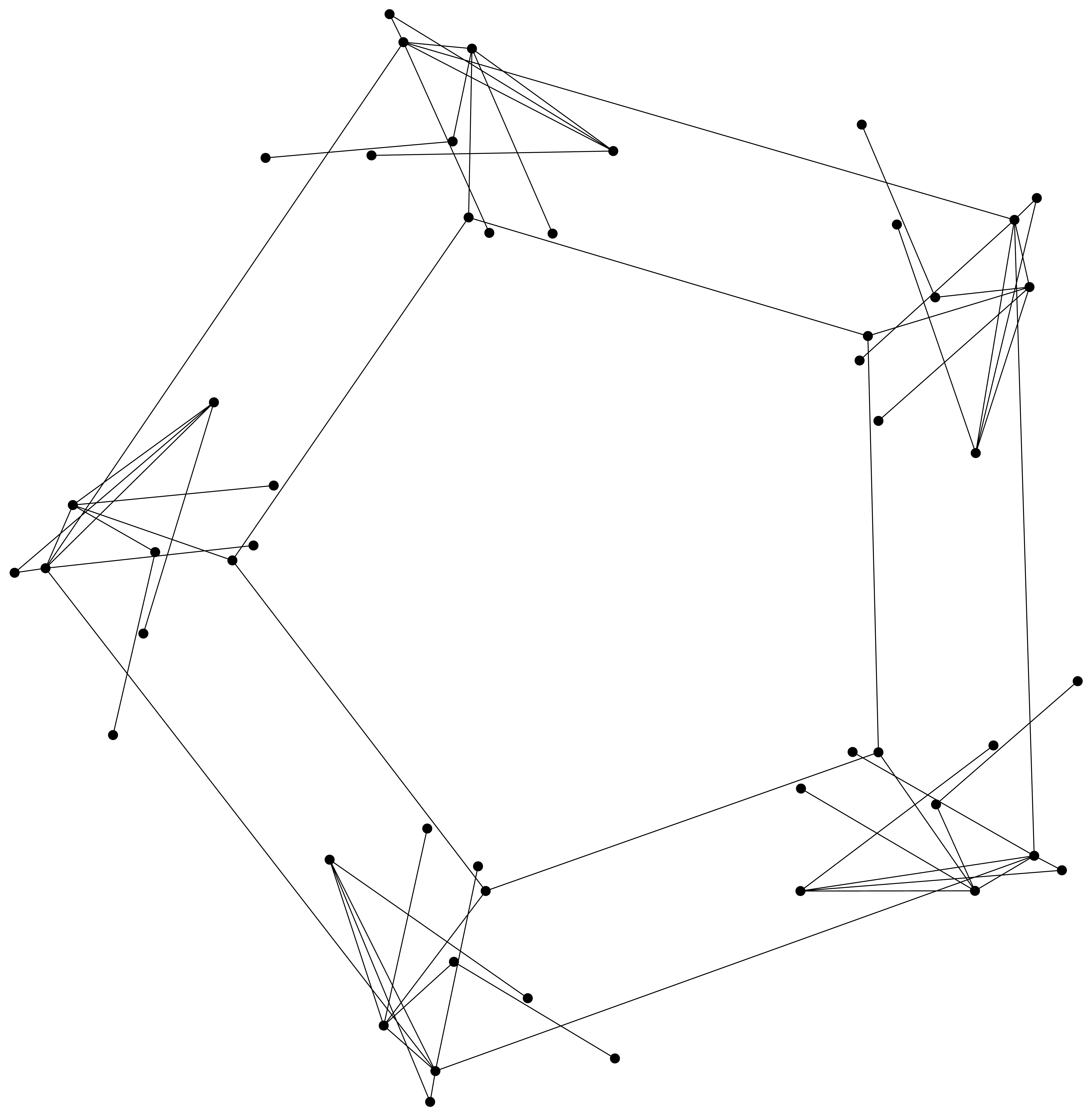}
                \caption{ }
                \label{fig:layoutRV5}
        \end{subfigure}%
        \begin{subfigure}[b]{0.14\textwidth}
                \includegraphics[width=\linewidth]{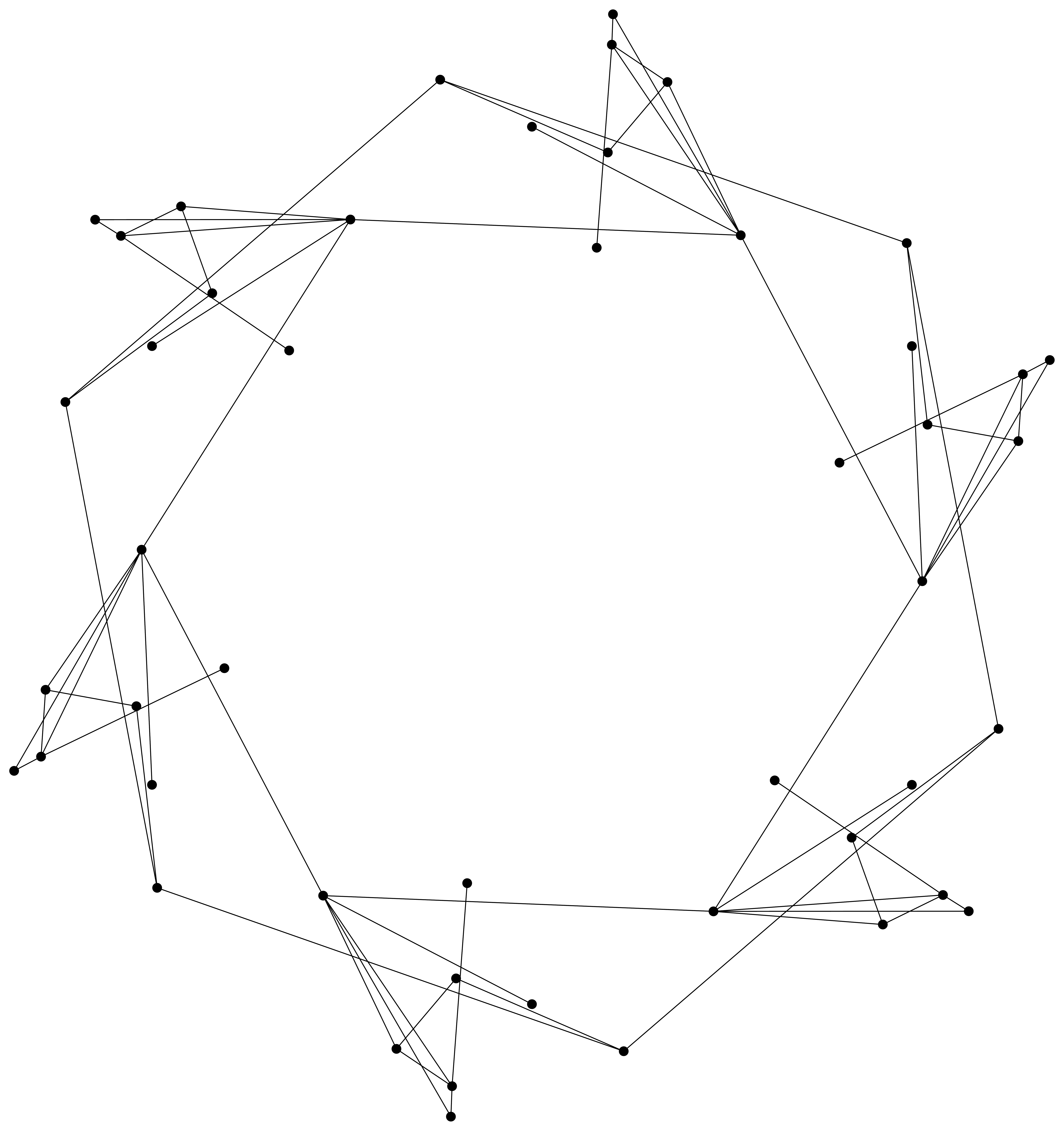}
                \caption{ }
                \label{fig:layoutRV6}
        \end{subfigure}%
        \begin{subfigure}[b]{0.14\textwidth}
                \includegraphics[width=\linewidth]{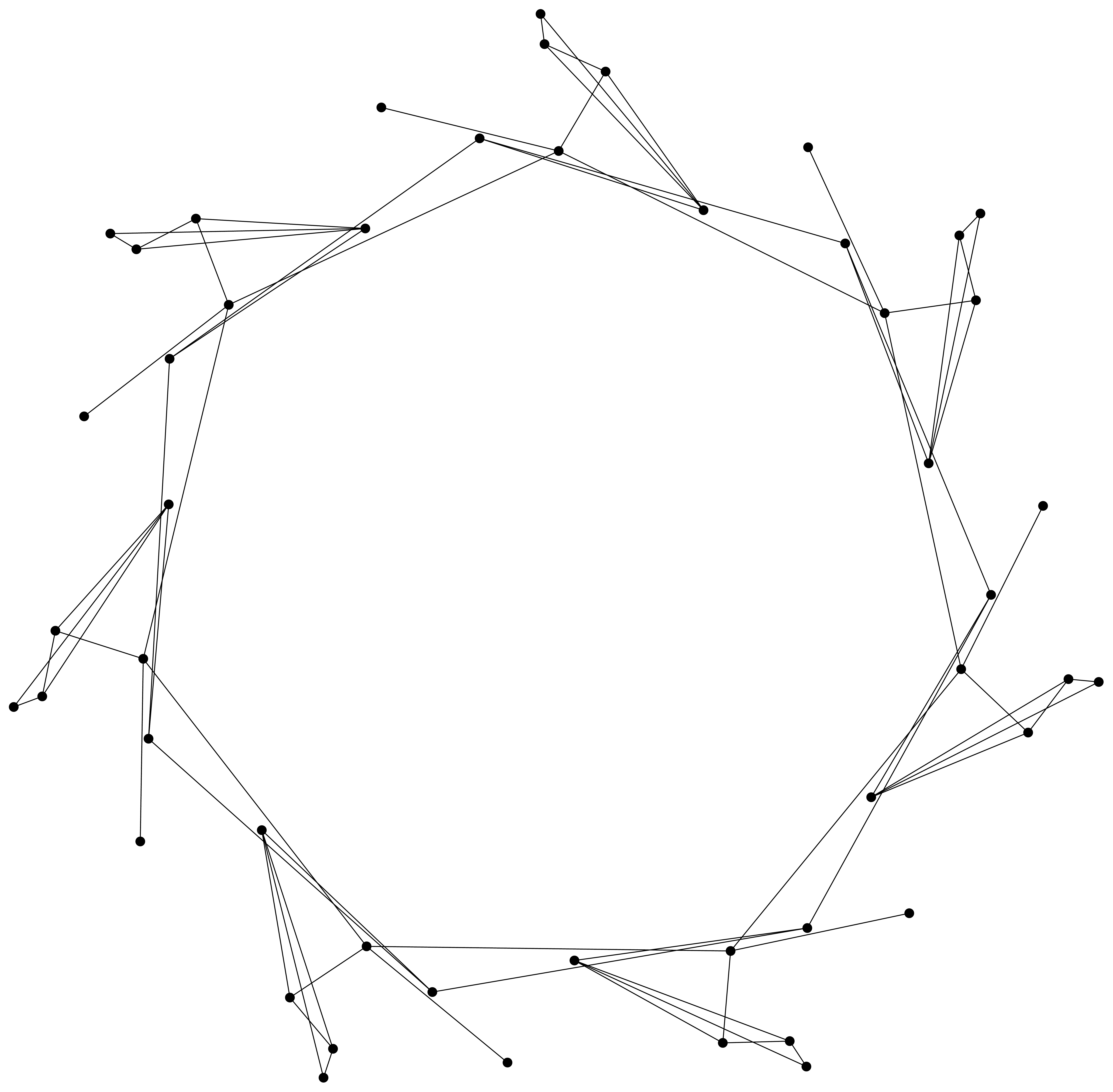}
                \caption{ }
                \label{fig:layoutRV7}
        \end{subfigure}%
        \begin{subfigure}[b]{0.14\textwidth}
                \includegraphics[width=\linewidth]{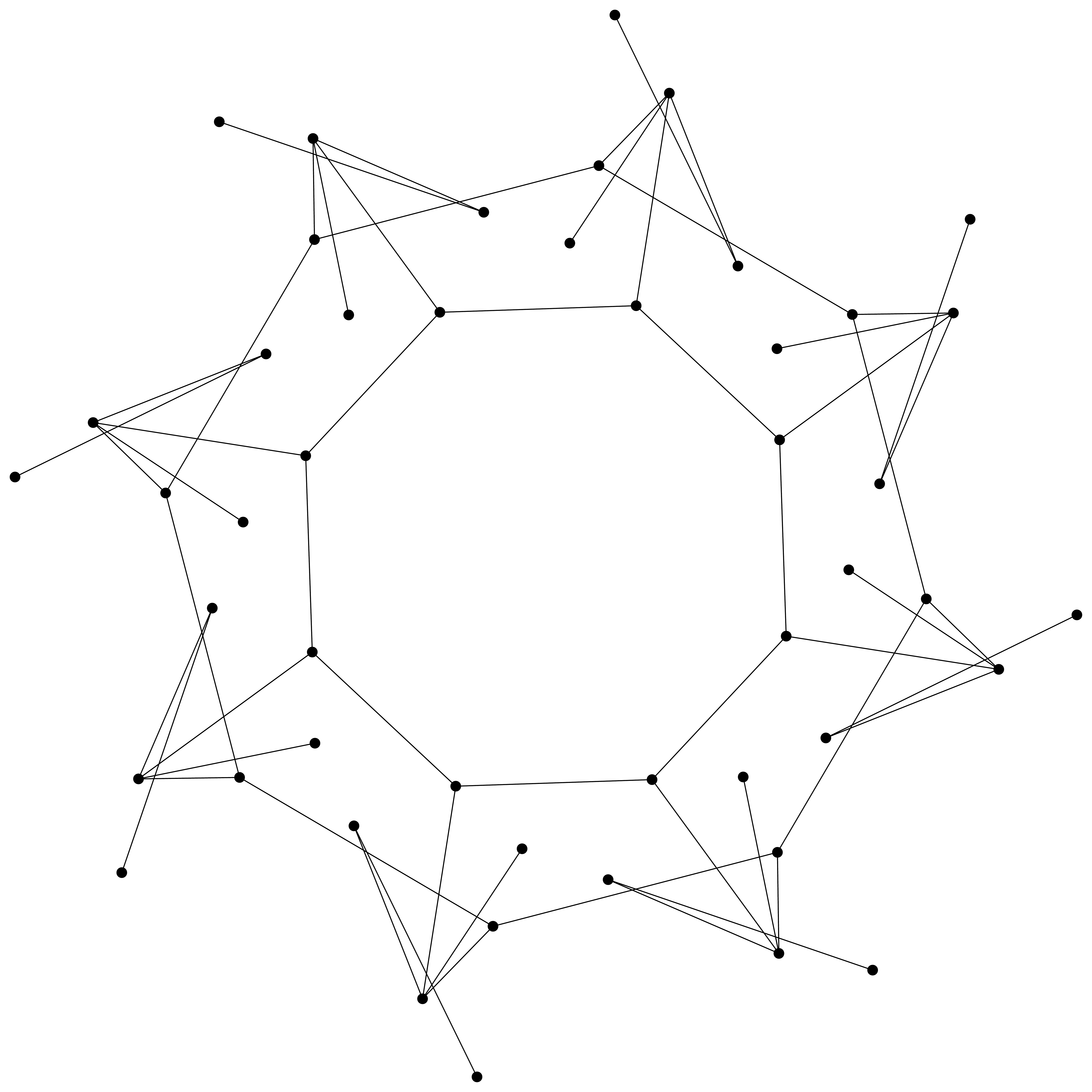}
                \caption{ }
                \label{fig:layoutRV8}
        \end{subfigure}%
        \begin{subfigure}[b]{0.14\textwidth}
                \includegraphics[width=\linewidth]{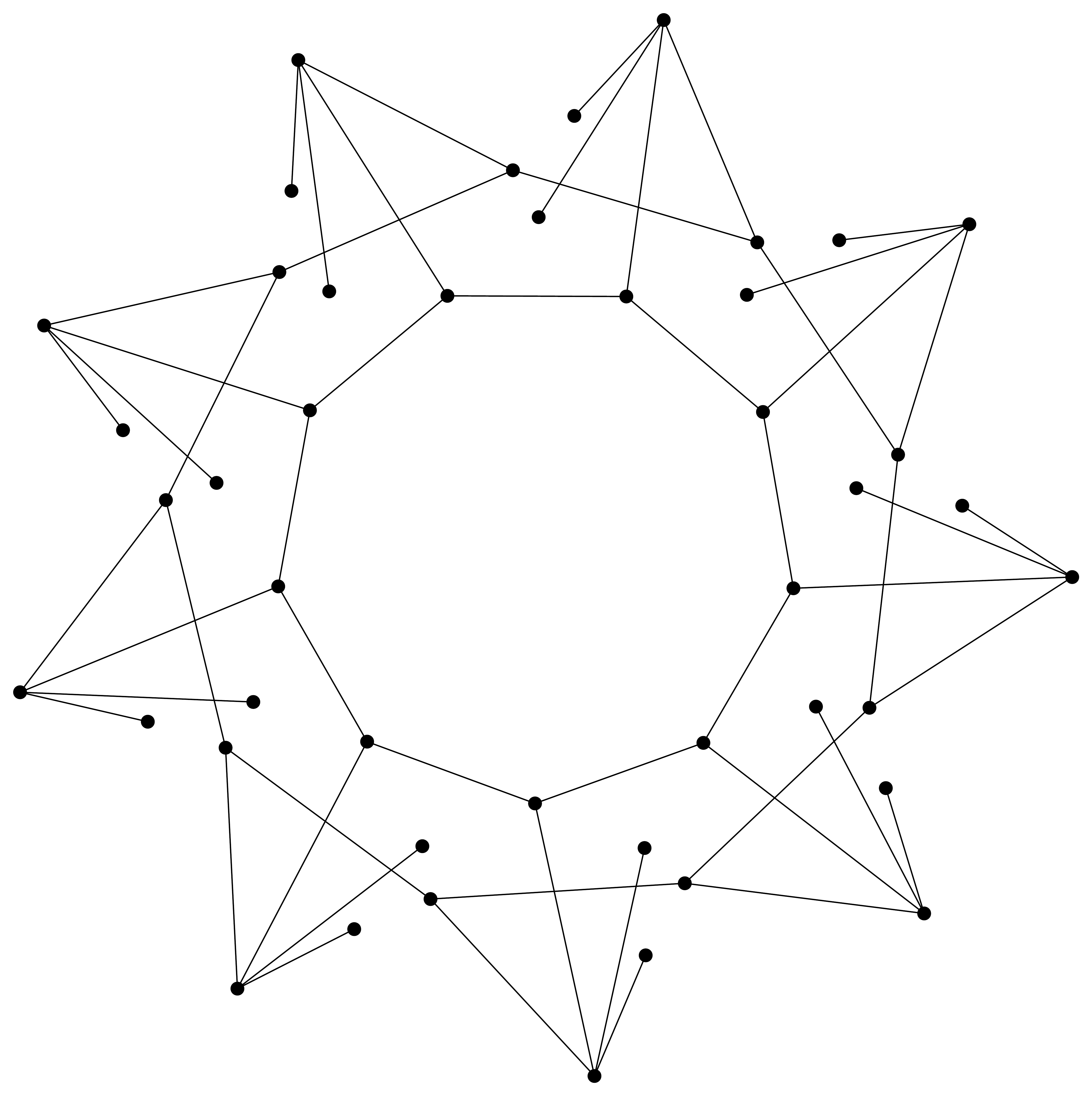}
                \caption{ }
                \label{fig:layoutRV9}
        \end{subfigure}%
        \begin{subfigure}[b]{0.14\textwidth}
                \includegraphics[width=\linewidth]{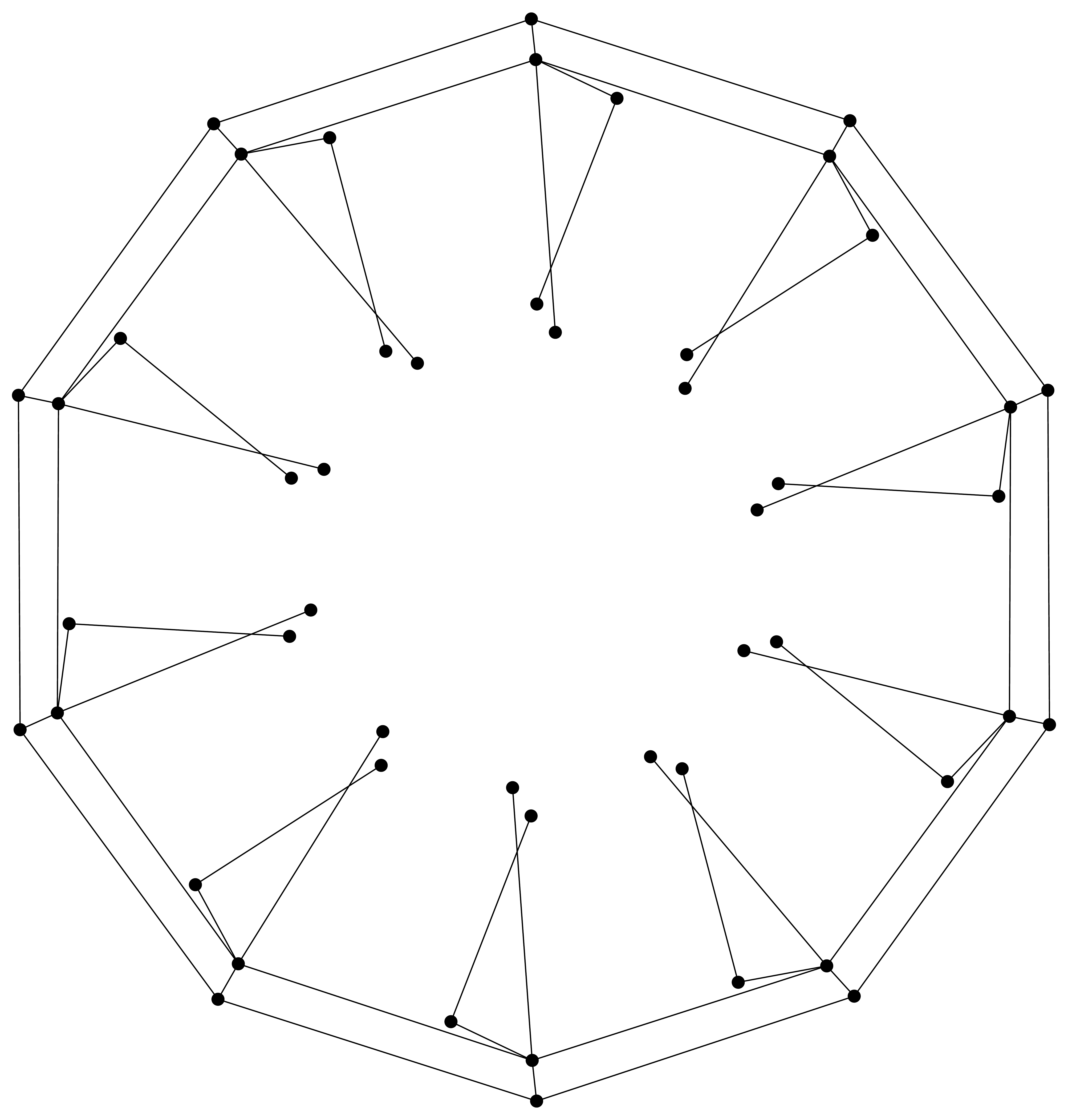}
                \caption{ }
                \label{fig:layoutRV10}
        \end{subfigure}%
        \caption{Example of rotational layout with fixed number of vertices: (a) RV4, (b) RV5, (c) RV6, (d) RV7, (e) RV8, (f) RV9, and (g) RV10.}\label{fig:layoutsradV}
\end{figure}

The rotational fixed vertices  (RV for short)  versions are symmetric layouts with a limited maximum number of total vertices. They are created as follows: 
\begin{itemize}
    \item \textbf{RV\{X\}}: 
    The base graph is reduced in size by removing as many vertices (at random) as needed so that when it is replicated on each of the $X = [4, \dots, 10]$ axes of symmetry, the total number of vertices does not exceed $50$.  As before, we connect each pair of rotationally consecutive components with two edges; see Fig.~\ref{fig:layoutsradV}. 
   Inevitably, the graphs in this set are not exactly of the same size, but they are within 20\% of each other; Table~\ref{tab:fixvertices} shows the number of vertices and average number of edges.
   \end{itemize}

We use small and sparse base graphs ($|V|=10$ and $|E|=11$) as these would be copied and interconnected when creating the various experimental sets. 
Each base graph is drawn by placing the vertices at random, to avoid accidental symmetries within the base graph and in order to focus on the symmetries created by the base graph replication. 
We use two (RC and RV) or three (HVT) edges to connect the replicated components for the same reasons.

By construction, all pairs of HVT graphs used as stimuli are the same, while pairs of RC and RV graphs are structurally different.
We used the yFiles library~\cite{yworks2005yed} 
to generate and draw our stimuli. 

\section{Experimental Methodology}
We conducted three separate experiments: Reflective and Translational (RT), Rotational Fixed Component (RFC) and Rotational Fixed Vertices (RFV). We used the same methodology but different sets of stimuli. The participants' task was to look at a series of pairs of graph drawings, and, for each pair, indicate which one of them they thought was ``more symmetric." This task was designed to address the following experimental questions:
\begin{itemize}
    \item \textbf{Q1}: Which type of symmetry among H, V, T, Hr, Vr, Tr is most recognizable as symmetry?
    \item \textbf{Q2}: How many rotations is most recognizable as rotational symmetry, using the fixed-component generation method?
    \item \textbf{Q3}: How many rotations is most recognizable as rotational symmetry, using the fixed-vertices generation method?
\end{itemize}

For Q1, we expected vertical symmetry (V) will be most prominent
, with horizontal (H) being noticed more than translational (T) 
, and that non-rotated versions will be easier to detect than the corresponding rotated versions
~\cite{bruce1975violations, royer1981detection, wagermans1995detection}.

For Q2 and Q3 we expected that the higher the degree, the greater the extent of symmetry recognition~\cite{jennings2017searching}. We were unsure about the effect of the additional visual clutter inherent in the RC drawings (as the size of graph increases with the degree) but anticipated that both should follow the same trend.

\subsection{Stimuli}
Each experiment uses five base graphs. Specifically, we generated 20 random simple graphs, called base graphs (each with 10 vertices and 11 edges), and drew them using a random layout. We randomly chose five of these base graphs as generators for the stimuli in each experiment.

\emph{Reflective and Translational (RT)}: There are 7 conditions in this experiment: horizontal (H), vertical (V),  translational (T), horizontal rotated (Hr), vertical rotated (Vr), translational rotated (Tr), and non-symmetric (NS). In this experiment, all drawings are of the same graph; they are all based on the same base component. With 5 different versions of the base graph, we have $5*7 = 35$ stimuli for this experiment.
 
\emph{Rotational Fixed Component (RFC)}: There are 7 conditions in this experiment: rotational orders of $4-10$ (designated as RC4, $\dots$, RC10). Since the number of incidences of the base component drawing is increased for each order, and two edges are added every time a new base component is added, the size of the graph increases with every order. With 5 different base graphs, we have $5*7 = 35$ stimuli for this experiment.
 
\emph{Rotational Fixed Vertices (RFV)}: There are 7 conditions in this experiment: rotational orders of $4-10$ (designated as RV4, $\dots$, RV10). Here, the number of vertices in each graph varies from 40 to 50, as described in 
Table~\ref{tab:fixvertices}. With 5 different versions of the base graph, we have $5*7 = 35$ stimuli for this experiment.

\section{Experimental Process}
We use a ``two-alternative forced choice" methodology, where a pair of stimuli are presented and participants must choose one of them. Specifically, the participants are asked to select the drawing that they considered ``more symmetric." In each experiment, we show all possible pairs twice (switching between left and right for the second presentation); with 7 stimuli we get $2 * (7 * 6)/2 = 42$ pairs and this is done for each of the 5 different versions of the base graph for a total of $5 * 42 = 210$ trials.
As a within-participants' experiment, all participants see all 210 pairs. To mitigate against the learning effect, the experimental stimuli are preceded by 10 practice trials for which the data is not collected (using a sixth version of the base graph), and each participant is presented with the 210 trials in a different random order. We collected data on the choice made by the participant for each pair, and the time taken to make the choice.
 
The experiment is conducted online 
with the online system randomly assigning one of the three experiments, and randomly selecting 5 base graphs from the 20 available for that participant-experiment combination. By asking each participant to do only one of the three experiments and by choosing only five versions of the base graph, we anticipated that the experiment would not be too lengthy (therefore minimizing the chance of participants not finishing the experiment). We expected the experiment to take approximately 10 minutes.

Participants are required to give consent at the start, and a set of instructions followed before the practice trials began. A self-timed break is offered every 20 trials.  
At the end of the experiment, participants are asked to give demographic data: gender, age, educational background, familiarity with networks and with symmetries. As a reward for taking part of the study, statistics about the participant's answers in relation to the answers of other participants are presented. These statistics show an example of the layout used in the task, the number of selections from the participants and from the specific participant grouped by each version, the number of clicks for any pair of versions, the number of left and right clicks, 
and the average answer time for the current and all participants.

Participants are not given details about the problem we are considering or the task that they would perform and we are  intentionally vague: we just ask them to \emph{``choose the layout that looks more symmetric"} between the proposed pair of layouts. We do not provide information about the concept of symmetry or our interpretations thereof, so as to not accidentally influence the participants.

\subsection{Pilot Study}
We use a simple online system made of four main parts: an  introduction page, the main experiment, a demographics page (gender, age, graph expertise, symmetry expertise), and the statistics page.
We conducted a pilot study to test our setup. 
In its early stages our system had an introductory screen with a brief description about the concept of \emph{``graphs"} and \emph{``graph drawings"} 
but several participants suggested to use \emph{``network"} instead, as it is more common and easier to understand. We also added information about the expected duration of the experiment. 
Initially we showed pairs of graphs starting from 10 base graphs for a total of 420 pairs, but the pilot study showed that this resulted in a test that was too long. We decided to reduce the number of base graphs to 5 and to introduce a break every 20 pairs. 
The demographics page was augmented to allow for feedback from the participants. 
The end of the study was modified from a ``Thank you" page to a page that provides statistics about the participants results in comparison to the average previous participants. 
Finally, following feedback from the pilot study, we changed the background color from white to light-gray 
and  added a black border to highlight the selected image.

\section{Data Analysis}
We used Reddit~\cite{reddit} and personal communications to crowdsource our study.
We collected data from a total of 97 participants.

\subsection{Experimental Conduct}
\begin{figure}
	\centering
	\begin{subfigure}[b]{0.24\textwidth}
		\includegraphics[width=\linewidth]{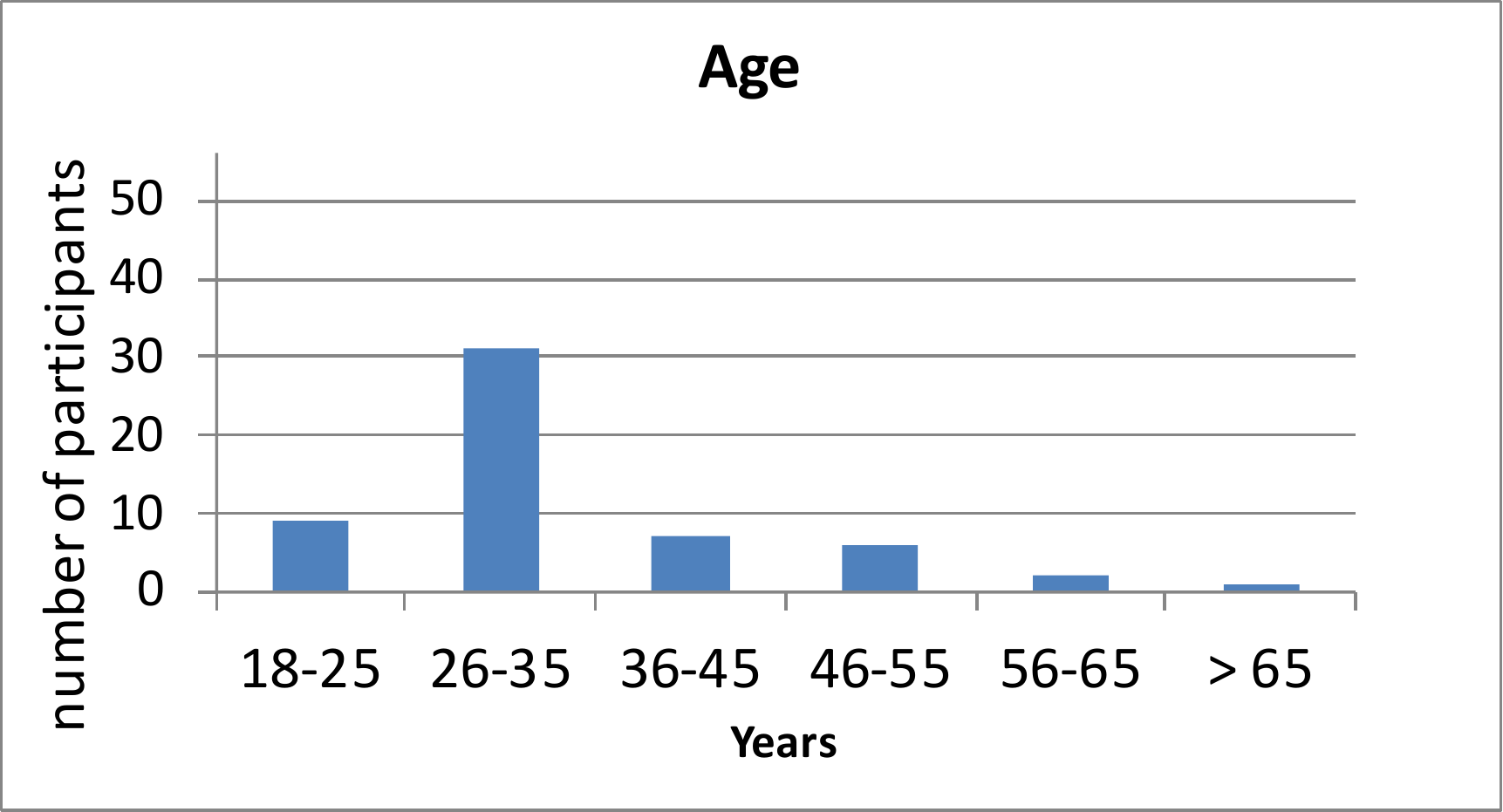}
		\caption{ }
		\label{fig:agechart}
	\end{subfigure}
	\begin{subfigure}[b]{0.24\textwidth}
		\includegraphics[width=\linewidth]{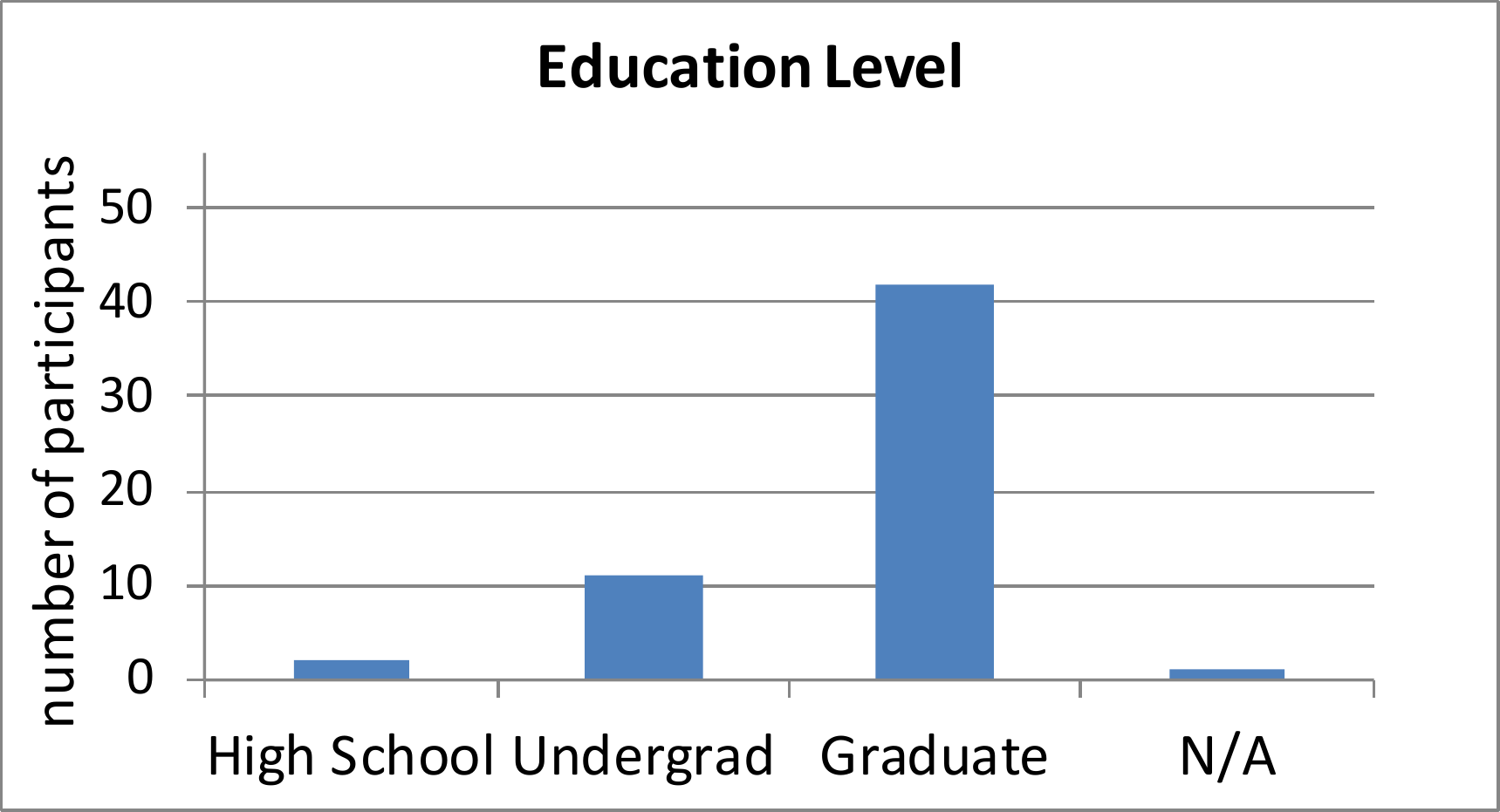}
		\caption{ }
		\label{fig:educhart}
	\end{subfigure}
	\begin{subfigure}[b]{0.24\textwidth}
		\includegraphics[width=\linewidth]{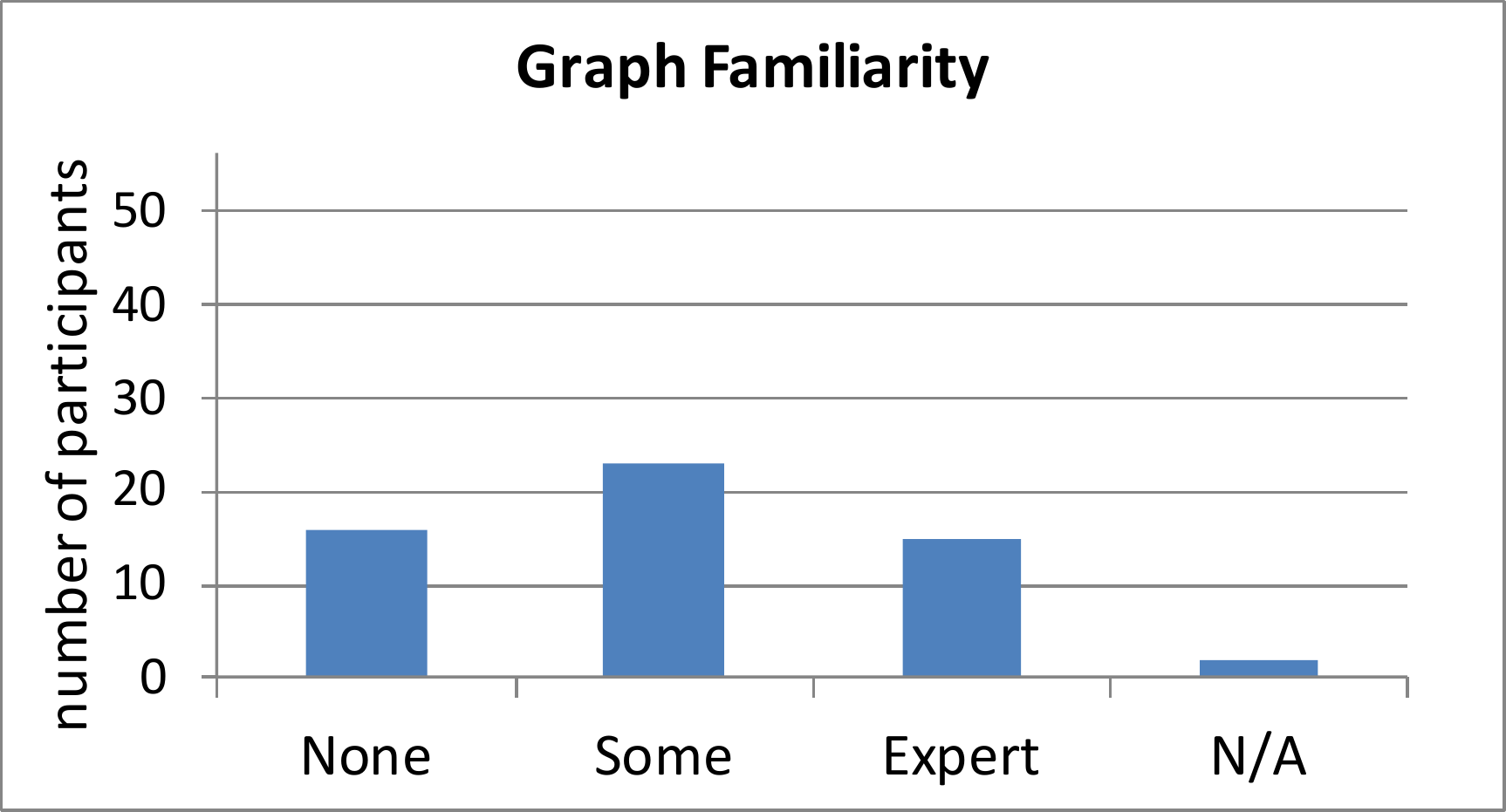}
		\caption{ }
		\label{fig:graphschart}
	\end{subfigure}
	\begin{subfigure}[b]{0.24\textwidth}
		\includegraphics[width=\linewidth]{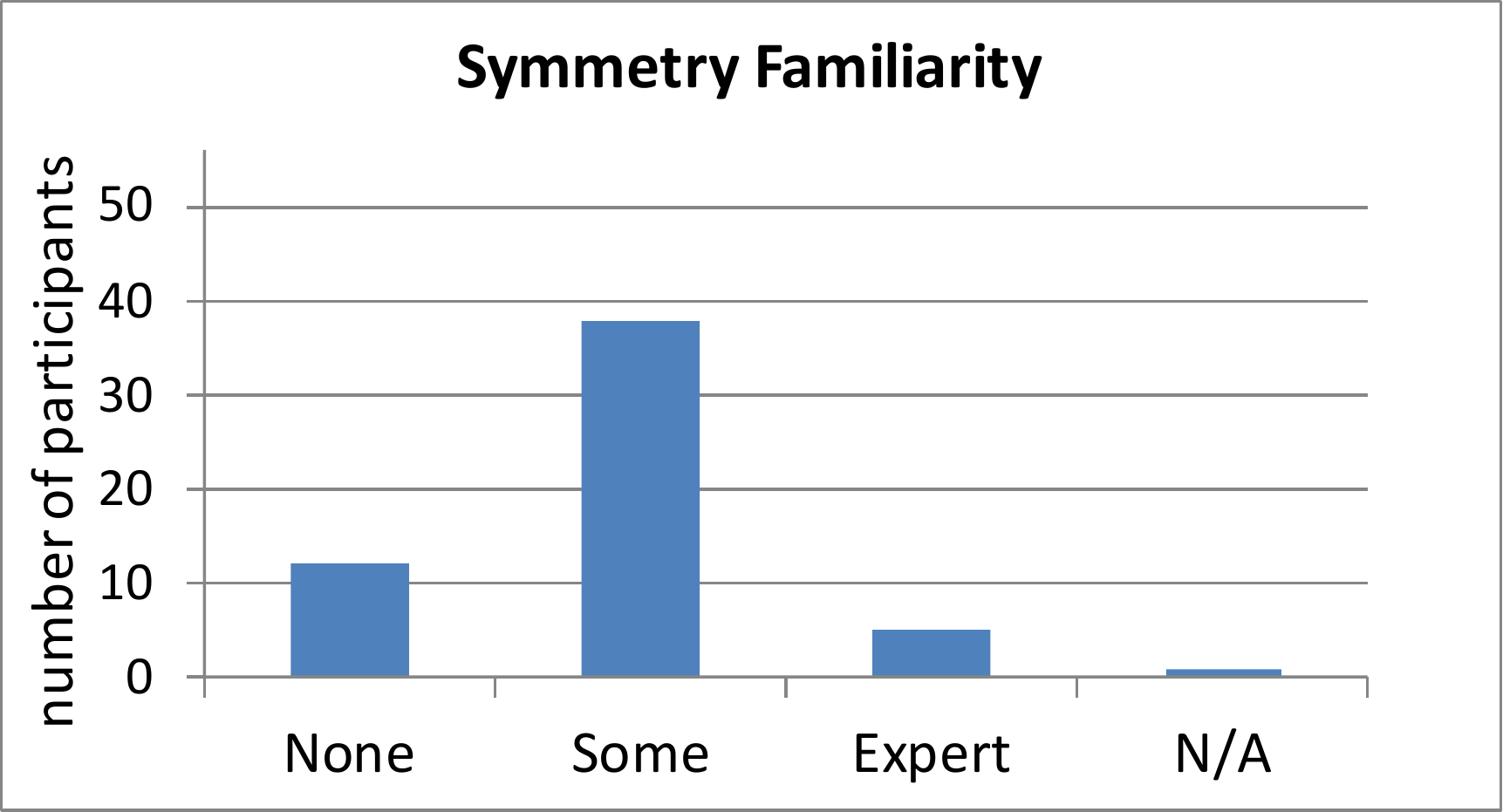}
		\caption{ }
		\label{fig:symchart}
	\end{subfigure}
	\caption{Participants demographic information (a) age, (b) education level, (c) graph familiarity, and (d) symmetry familiarity.}
	\label{fig:democharts}
\end{figure}

We removed data from all participants who did not complete the full experiment ($n=39$). It's disappointing that so many people did not finish experiment (which was intended to take less than ten minutes); informal feedback from participants are that they found the task boring and, in some cases, difficult.
One participant in the RT experiment gave an exceptionally high number of votes to the non-symmetric drawing; we removed this participant from our analysis.
Two response times were particularly higher than others ($17s$ and $100s$) - these data points were replaced by the mean of the other response times for the respective participants. One participant gave almost exactly the same votes to all conditions in the RFV experiment: we removed this participant's data. This left us with 19 participants for RT, 19 for RFC and 18 for RFV. The demographic information of the 56 participants ($13$ female and $43$ male) is summarized in Fig.~\ref{fig:democharts}.

\subsection{Analysis Process}

\begin{figure}
	\centering
	\begin{subfigure}[b]{0.31\textwidth}
		\includegraphics[width=\linewidth]{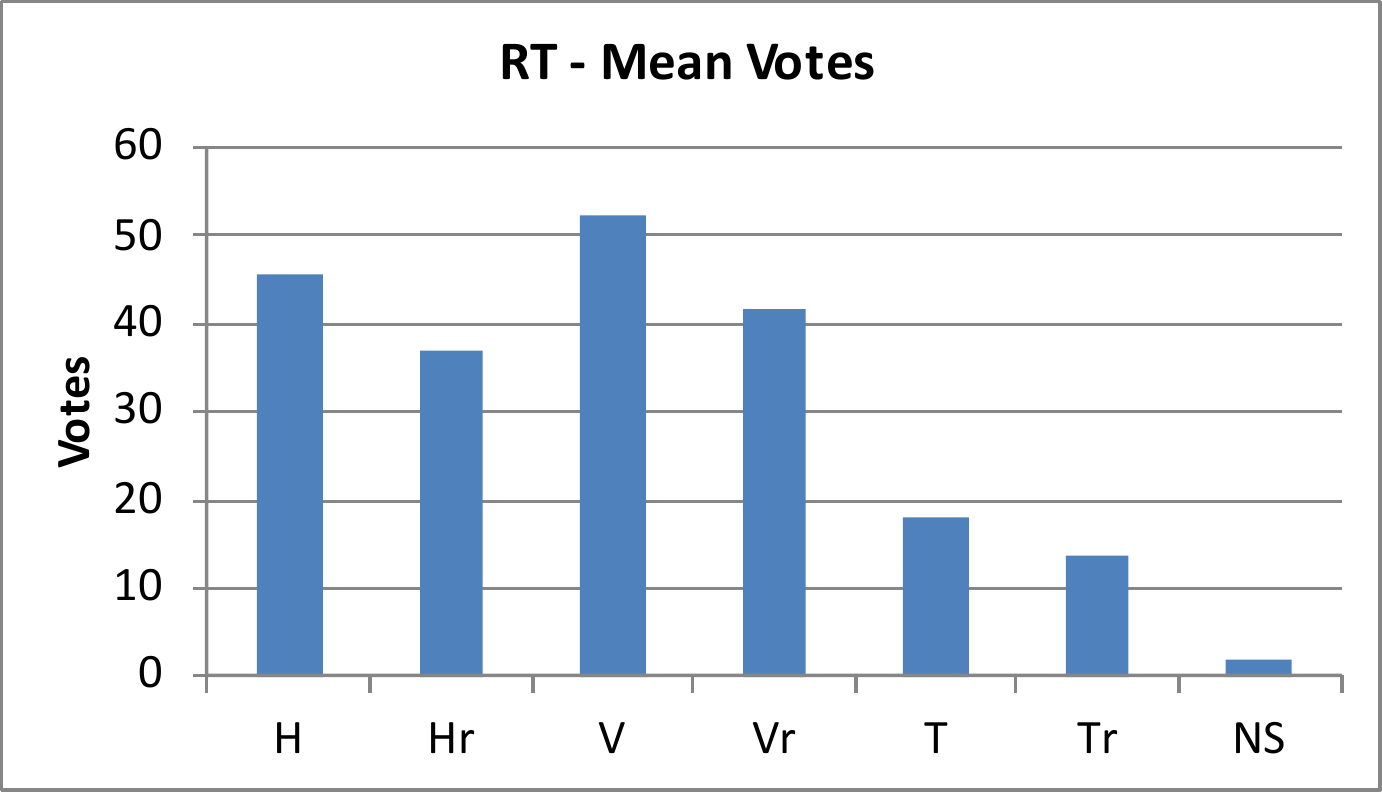}
		\caption{ }
		\label{fig:chartTotVotesHVT}
	\end{subfigure}
	\begin{subfigure}[b]{0.31\textwidth}
		\includegraphics[width=\linewidth]{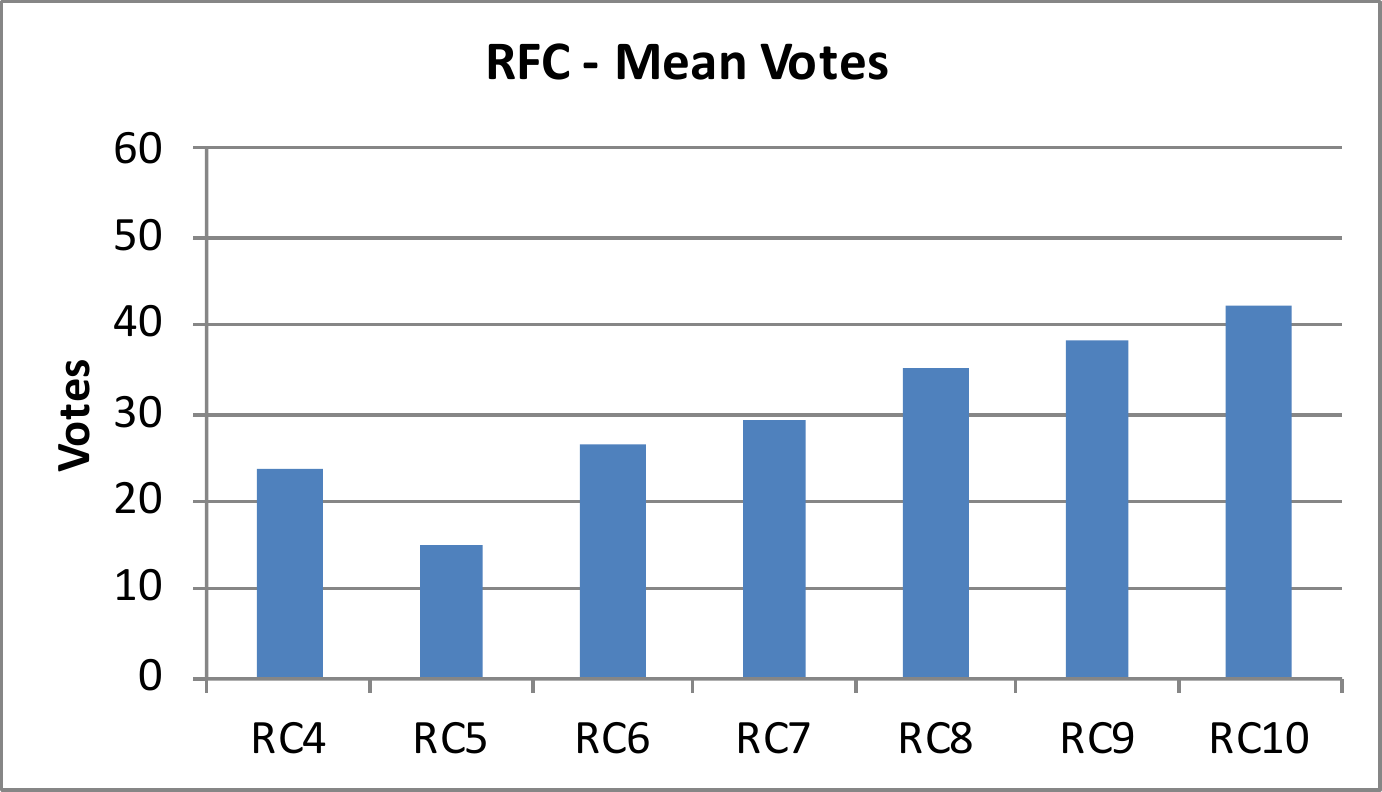}
		\caption{ }
		\label{fig:chartTotVotesRC}
	\end{subfigure}
	\begin{subfigure}[b]{0.31\textwidth}
		\includegraphics[width=\linewidth]{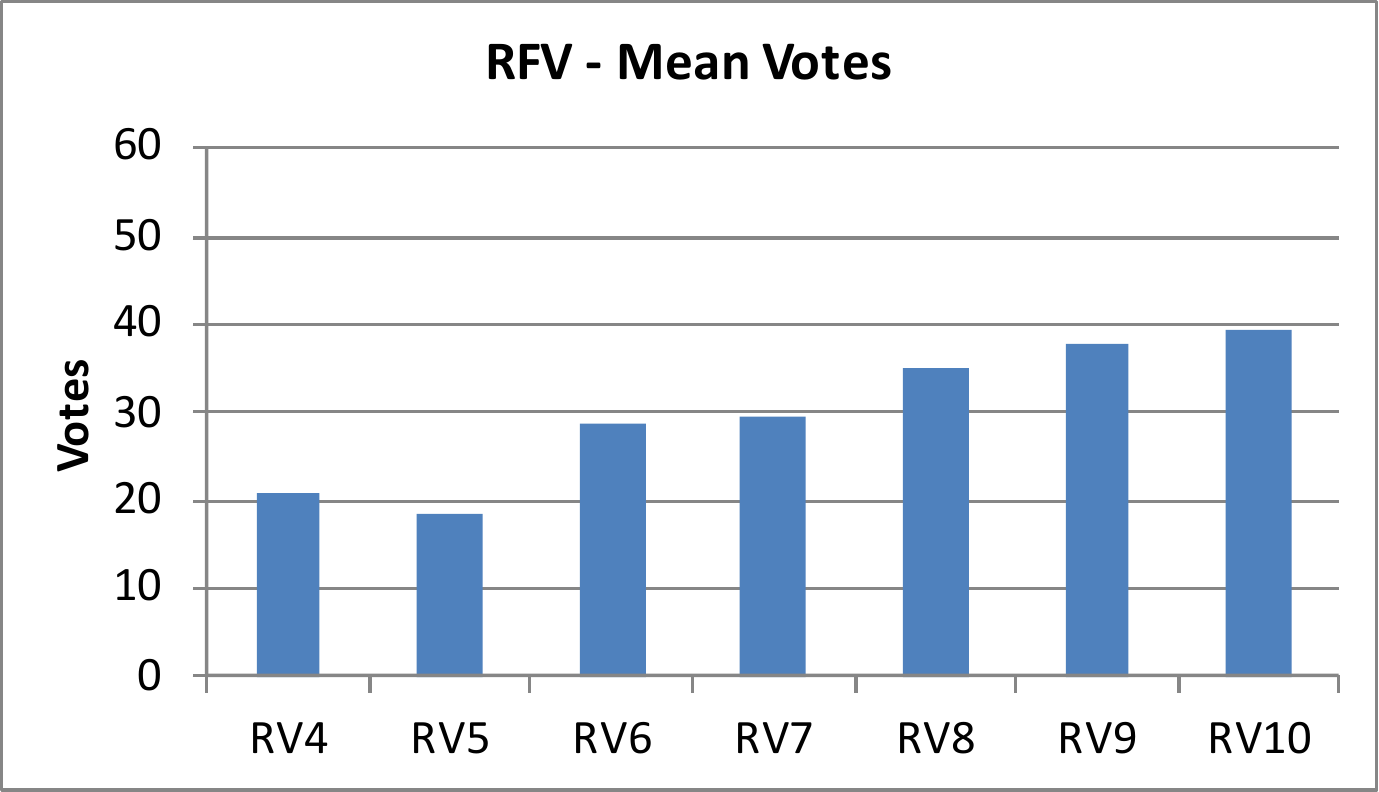}
		\caption{ }
		\label{fig:chartTotVotesRV}
	\end{subfigure}
	
	\begin{subfigure}[b]{0.31\textwidth}
		\includegraphics[width=\linewidth]{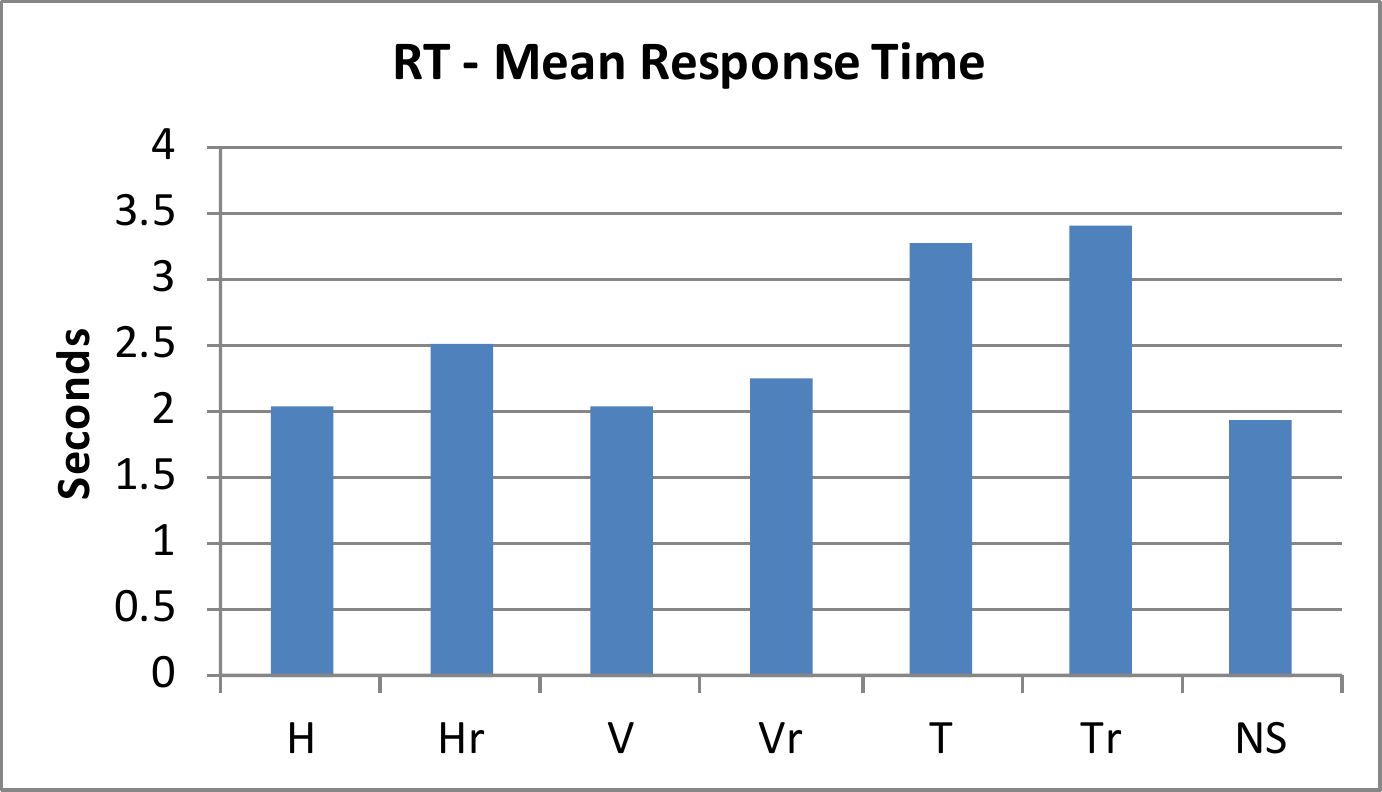}
		\caption{ }
		\label{fig:chartAVGTimeHVT}
	\end{subfigure}
	\begin{subfigure}[b]{0.31\textwidth}
		\includegraphics[width=\linewidth]{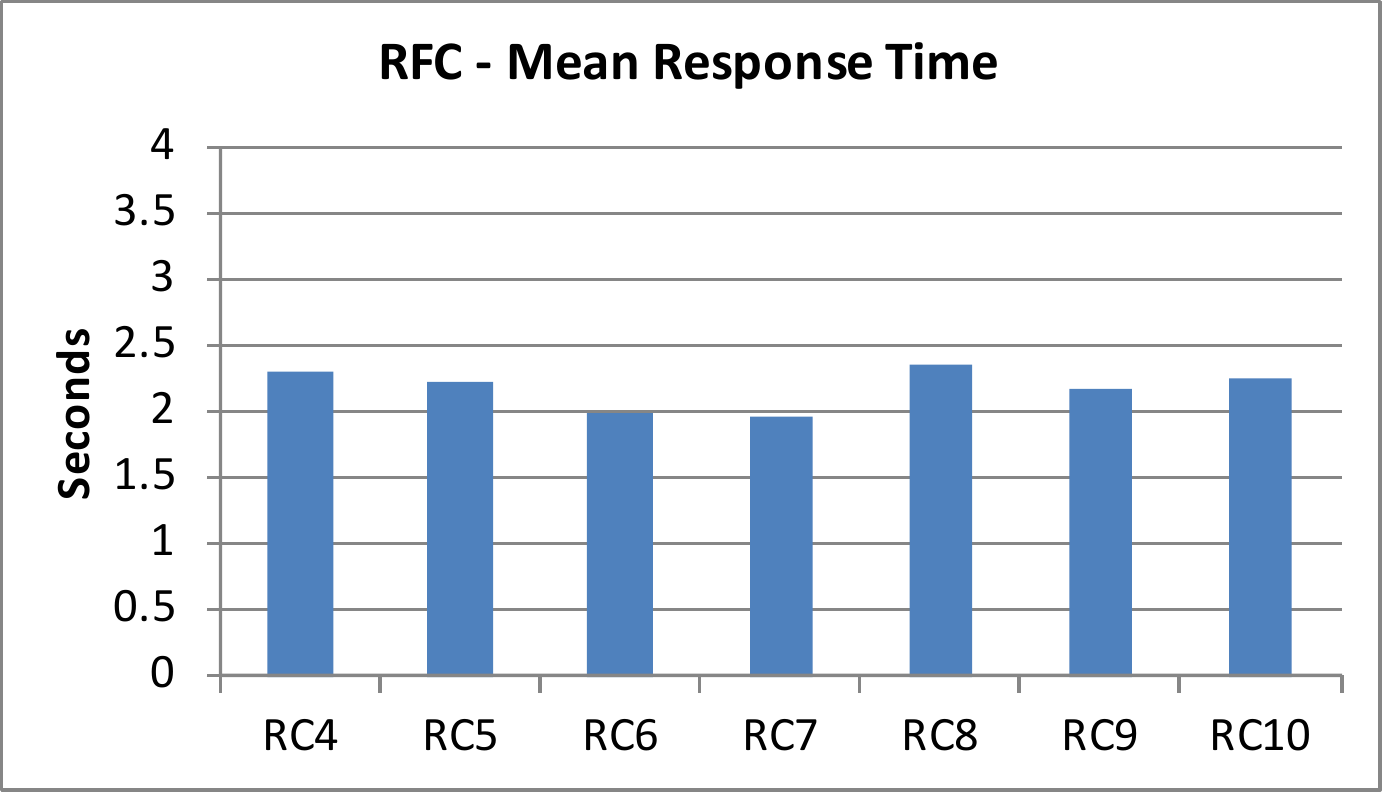}
		\caption{ }
		\label{fig:chartAVGTimeRC}
	\end{subfigure}
	\begin{subfigure}[b]{0.31\textwidth}
		\includegraphics[width=\linewidth]{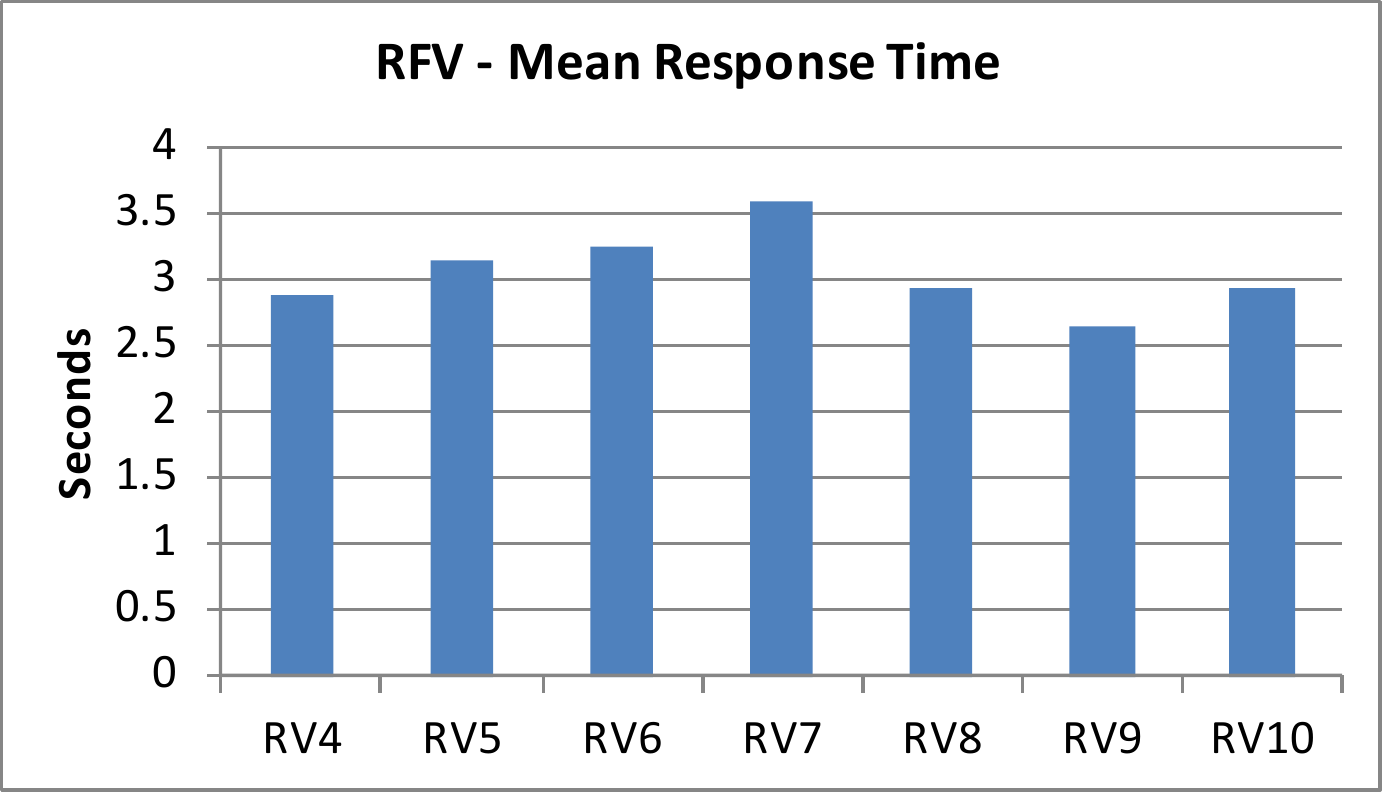}
		\caption{ }
		\label{fig:chartAVGTimeRV}
	\end{subfigure}
	\caption{Mean votes charts for (a) RT (b) RFC, (c) RFV and mean response time charts for (d) RT (e) RFC, and (f) RFV. }\label{fig:chartMeanTime}
	\label{fig:chartMeanVotes}
\end{figure}

We (conveniently) have seven conditions for each of the three experiments, so the form of the data is the same. Since we include only those participants who completed the entire experiment, each participant ``voted" 210 times, each vote being associated with one of the seven conditions. We also have an average response time associated with each vote. For both response times and votes, we use ANOVA and adjusted planned comparison pair-wise tests between the pairs of conditions of interests to determine which conditions are (a) favored over the others, and (b) responded to most quickly. We use a significance level of 0.05 throughout, adjusted as appropriate for the number of planned comparisons made. We do not compare all pairs of conditions, only those of interest to our research question as doing this reduces the extent of required adjustments.
The mean vote and mean response time (over all participants) charts are depicted in Fig.~\ref{fig:chartMeanVotes}, while the exact values are shown in Table~\ref{tab:meanvotetime}.

\paragraph{\textbf{Voting Responses}}
~\\
\textit{Reflective and Translational (RT):}
A repeated-measures ANOVA reveals a significant difference between the average votes ($F= 240.5$, $df= 6$  , $p<0.001$ ). Five pairwise comparisons (at adjusted $p=0.01$), reveal the following results;
\begin{itemize}
    \item H obtained significantly more votes than Hr ($p<0.001$)
    \item V obtained significantly more votes than Vr ($p<0.001$)
    \item T obtained significantly more votes than Tr ($p<0.001$)
    \item V obtained significantly more votes than H ($p<0.001$)
    \item H obtained significantly more votes than T ($p<0.001$)
\end{itemize}

\paragraph{Rotational Fixed Component (RFC):}
A repeated-measures ANOVA reveals a significant difference between the average votes ($F= 12.2$, $df= 6$, $p<0.001$). Two pairwise comparisons (at adjusted $p=0.025$), reveal the following results;
\begin{itemize}
    \item RC6 obtained significantly more votes than RC5 ($p<0.001$)
	\item No significant difference is between the votes for RC4 and RC5 ($p=0.050$)
\end{itemize}
 
\paragraph{Rotational Fixed Vertices (RFV):} 
A repeated-measures ANOVA reveals a significant difference between the average votes ($F= 10.9$, $df= 6$, $p<0.001$). Two pairwise comparisons (at adjusted $p=0.025$), reveal the following results;
\begin{itemize}
    \item RV6 obtained significantly more votes than RV5 ($p=0.021$)
    \item No significant difference is between the votes for RV4 and RV5 ($p=0.63$)
\end{itemize}

\begin{table}
	
	\begin{center}
		\caption{Mean vote and mean response time for each task.}
		\begin{tabular}{ | l | c | c | c | c | c | c | c | c |}
			\hline
			& \textbf{\begin{tabular}{@{}c@{}}number of \\ participants\end{tabular} } & \textbf{H} & \textbf{Hr} & \textbf{V} & \textbf{Vr} & \textbf{T} & \textbf{Tr} & \textbf{NS} \\ \hline
			\textbf{mean vote} & 19 & 45.63  & 36.89 & 52.16 & 41.74 & 18 & 13.63 & 1.95 \\ \hline 
			\textbf{mean response time (s)} & 19 & 2.05 & 2.53 & 2.03 & 2.25  &  3.27 & 3.41 &  1.93\\ \hline \hline
			& & \textbf{RC4} & \textbf{RC5} & \textbf{RC6} & \textbf{RC7} & \textbf{RC8} & \textbf{RC9} & \textbf{RC10} \\ \hline
			\textbf{mean vote} & 19 & 23.63 & 15.11 & 26.42 & 29.05 & 35.21 & 38.32 & 42.26 \\ \hline 
			\textbf{mean response time (s)} & 19 & 2.30 & 2.21 & 1.98 & 1.96 & 2.36 & 2.18 & 2.24 \\ \hline \hline
			& & \textbf{RV4} & \textbf{RV5} & \textbf{RV6} & \textbf{RV7} & \textbf{RV8} & \textbf{RV9} & \textbf{RV10} \\ \hline
			\textbf{mean vote} & 18 & 20.78 & 18.67 & 28.78 & 29.44 & 35.22 & 37.72 & 39.39 \\ \hline 
			\textbf{mean response time (s)} & 18 & 2.89 & 3.15 & 3.25 & 3.58 & 2.94 & 2.65 & 2.93 \\ \hline 
		\end{tabular}
		\label{tab:meanvotetime}
	\end{center}
\end{table}

\paragraph{\textbf{Response Time}}
~\\
\textit{Reflective and Translational (RT):}
A repeated-measures ANOVA reveals a significant difference between the average response time ($F= 4.84$, $df= 6$, $p<0.001$). Five pairwise comparisons (at adjusted $p=0.01$), reveal;
\begin{itemize}
    \item The response time for H is significantly less than that for T ($p<0.001$)
    \item No significant differences between the response times for the following pairings: H/Hr; V/Vr; T/Tr; V/H.
\end{itemize}

\paragraph{Rotational Fixed Component (RFC):}
A repeated-measures ANOVA reveals no significant difference between the average response time ($F= 1.29$, $df= 6$, $p=0.267$). 
 
\paragraph{Rotational Fixed Vertices (RFV):}
A repeated-measures ANOVA reveals no significant difference between the average response time ($F= 1.535$, $df=6$, $p=0.174$).

\section{Discussion}

With respect to question \textbf{Q1}, we found statistically significant effects confirming 
that vertical symmetry is more recognizable as symmetry, followed by horizontal and translational.
In all three variants adding even slight rotation has a significant effect. Vertical symmetry is more recognizable than vertical with rotation, horizontal is more recognizable than horizontal with rotation, and translational is more recognizable than translational with rotation.
Mean response time also follows this trend, although we have only one statistically significant finding that horizontal is faster than translational.
An immediate implication of this is that vertical symmetry is the best perceived because it is frequently seen by people and that can be exploited in future graph layout algorithms.

The RFC and RFV experiments helped us answer questions \textbf{Q2} and \textbf{Q3}. Our experimental results provide evidence of a greater symmetry recognition for high number of rotation axes. 
We believe that this is due to the fact that  layouts with high number of rotational axes tend to become more and more  circular, and the circle is a very symmetric shape.
We also find that the increased size of the graphs in the RFC experiments does not seem to affect the better perception for high number of rotational axes as the results for RFC and RFV are similar.

Of particular interest is one exception: RC4 is considered more symmetric than the RC5, which goes against the general trend of better perception for high number of rotation axes. We discuss possible explanations below.

\section{Conclusions and Future Work}\label{sec:conclusions}
The conclusions from our experiments are limited by the specifics of our study; e.g.,  we only consider graphs of similar sizes, starting with a base graph of a fixed size, and connecting copies thereof in symmetric sub-structures using two or three edges. Despite such limitations, our experiment does provide potentially useful information about the relative effects of different types of symmetries in drawings of graphs.
The results from our study suggest that humans recognize vertical reflective symmetry over all other types of symmetry, followed by horizontal and translational symmetries and that rotational symmetry is affected by the number of radial axes.
These findings can help guide algorithms that identify features to be displayed using these types of symmetries. Vertical symmetry can be used to call attention to isomorphic  subgraphs and cycles can be highlighted by laying them out as regular $n$-gons that have high rotational symmetry.

The results of our experiment, in particular for the RFC  and RFV tasks, show that a rotationally symmetric layout with 4 axes is considered more symmetric than that with 5 axes, which goes against the general tendency of recognizability  towards higher number of axes. This leads to an interesting question: 
what is it about these rotationally symmetric layouts that gives such different results? Is it because of something in particular about the layout with 5 axes or the one with 4? Is it because rotational 4 is not perceived as a rotational symmetry but as a combination of horizontal and vertical symmetries? If so, is the rotational symmetry with many axes perceived better than the reflective and translational ones?
Answers to such questions can help guide the design of algorithms that visualize symmetries. 
It would also be interesting to repeat this study using different base graphs that were not drawn with random layout.

\section*{Acknowledgements} We thank Eric Welch for help with the experiment, Hang Chen for help with statistics, and yFiles for providing their software for research and education. This work is supported by NSF grants CCF-1423411 and CCF-1712119.

 \bibliographystyle{splncs03}

\end{document}